\documentclass[oneside]{article}
\usepackage{amsmath, amssymb, amsfonts}
\usepackage{epsfig}
\begin{document}
\newenvironment{ttabbing}{\begin {tabbing}
                                \tt }{\end{tabbing}}
\newcommand {\bet} {\scriptsize \begin{ttabbing}}
\newcommand {\ent} {\end {ttabbing}}
\newcommand {\enum} {\begin{enumerate}}
\newcommand {\enume} {\end{enumerate}}
\newtheorem{example}{Example}[section]
\newtheorem{sublemma}[example]{Sublemma}
\newtheorem{theorem}[example]{Theorem}
\newtheorem{lemma}[example]{Lemma}
\newtheorem{definition}[example]{Definition}
\newtheorem{proposition}[example]{Proposition}
\newtheorem{conjecture}[example]{Conjecture}
\newtheorem{corollary}[example]{Corollary}
\newtheorem{remark}[example]{Remark}
\normalsize              
\begin{center}
{\Large Dynamics of a Ring of Diffusively Coupled Lorenz Oscillators} \\
\vspace{1cm}
Kre\v{s}imir Josi\'{c} \\ Department of Mathematics,
Pennsylvania State University \\ present address: Department of
Mathematics, Boston University \\ 111 Cummington Street, Boston, MA
02215 \\ \vspace{1cm}
C. Eugene Wayne \\ Department of
Mathematics, Boston Universty \\ 111 Cummington Street, Boston, MA
02215 \\ \vspace{1cm} 
\end{center}

\begin{abstract} We study the dynamics of a finite chain of
diffusively coupled Lorenz oscillators with periodic boundary
conditions.  Such rings possess infinitely many fixed states, some of
which are observed to be stable.  It is shown that there exists a
stable fixed state in arbitrarily large rings for a fixed coupling
strength.  This suggests that coherent behavior in networks of
diffusively coupled systems may appear at a coupling strength
that is independent of the size of the network.   
\end{abstract}

\section{Introduction}

Lattices of coupled dynamical systems have been studied in many
different contexts: as discrete
versions of partial differential equations of evolution type
\cite{BuSi}, \cite{Radu}, models of neuronal 
networks \cite{aron}, \cite{huerta} and phase lock loops 
\cite{nekor}, and in statistical mechanics \cite{fpu}.  The
particular problem of synchronization and emergence of coherent behavior in
lattices of diffusively coupled, 
chaotic, continuous time dynamical systems has received
much attention recently due to its applications in neuroscience \cite{abarbanel},
\cite{huerta}, and chaotic synchronization \cite{josic}, \cite{heagy}.

Analytical results in the literature suggest that the coupling
strength necessary for the appearance of coherent behavior in networks
of chaotic systems grow with the size of the network
\cite{lin},\cite{nekor},\cite{hale}.  For instance,  
estimates of Afraimovich and Lin \cite{lin} suggest that the coupling
strength necessary to synchronize a lattice of
generalized forced Duffing systems grows as a high power of the size
of the lattice.  Since stable, coherent states in lattices may
frequently be viewed as instances of generalized synchronization
\cite{josic}, one might guess that the appearance of dynamically
stable states of this type also requires that the coupling grows
with the number of oscillators in the system.  
This would make it unlikely that such networks
are of importance in nature, since the coupling strength 
in realistic systems cannot be 
varied by orders of magnitude.

In contrast to these theoretical results, the 
numerical experiments of Huerta, et. al. \cite{huerta} indicate that
partial synchronization may appear in certain networks of chaotic
systems at a nearly constant coupling strength, regardless of the size
of the network. In particular, \cite{huerta} considered a system
with about $10^4$ coupled neurons.  Each neuron was modeled
by a system of three nonlinear, ordinary differential equations which,
when uncoupled, underwent chaotic motion.  They observed a variety
of different stable, coherent motions including a number
of states that exhibited roughly periodic patterns.
 It is the goal of this paper to show rigorously that
such such stable, coherent states can occur in arbitrarily large
systems of coupled chaotic oscillators at a coupling strength
which is independent of the size of the system.

Since the Lorenz system is  one of the paradigms of
chaotic flows, we have chosen a ring of diffusively coupled Lorenz
equations for our investigations.  The systems in the ring evolve
according to the equations
\begin{eqnarray}\label{coupledlorenz}
x_i' = \sigma (y_i-x_i) + d_x \Delta x_i \nonumber \\
y_i' = r x_i - y_i - x_i z_i + d_y \Delta y_i \label{eq1} \\ 
z_i' = - \beta z_i + x_i y_i + d_z \Delta z_i \nonumber
\end{eqnarray}
where $\Delta x_i =  x_{i-1} - 2 x_i + x_{i+1}$ with the index $i$
taken modulo $n$ is the discretized Laplacian with periodic boundary
conditions.  The constants 
$\sigma = 10, r = 27$, and $\beta = 8/3$ are chosen so that 
each uncoupled system would be in the chaotic regime. Following the 
work of \cite{huerta}, we will focus on the case $d_x \neq 0$, 
$d_y = d_z = 0$.  
As shown in \cite{josic}, such a ring
is synchronized identically for sufficiently large
values of $d_x$. We will be concerned with the case when $d_x$
is sufficiently large for coherent spatial structures to emerge in the
chain, but insufficient to synchronize the ring.  

The paper is organized as follows:  Typical spatial
structures observed in numerical experiments are described in section
\ref{numerics}. In particular, we give examples of stable stationary
states, traveling waves, and breathers. 
In section \ref{steadystate} a general framework for the
study of fixed states in chains and rings of dynamical systems is
discussed.  We show that the existence of such states can be
reduced to the study of periodic solutions of reversible dynamical
systems on ${\Bbb{R}}^n$.  We then apply these results to two different cases,
including the chains of coupled Lorenz oscillators (\ref{coupledlorenz}),  
to show that a large number of fixed states
can be expected in such rings.  The conditions under which these
states are stable are investigated in section \ref{stabilityofsteady}.
We  then show how Floquet theory  can be used to reduce the
stability question in lattices of arbitrarily many oscillators
to the study of the spectrum of a matrix of fixed size (but depending
on a parameter -- the ``quasi-momentum''.)
Finally, in section \ref{example} we  implement this program to
rigorously construct an example of a spatially periodic
stable fixed state which remains stable for rings of arbitrary size. 

\section{Numerical Experiments} \label{numerics}

Numerical experiments were performed on a Sun SuperSparc using the
numerical integration program XPP. Several different integration
methods were used to check the validity of the results. These included
 fixed and variable step Runge-Kutta methods of
varying order and the Gear method.  

At very low values of the coupling constant $d_x$, the individual systems
in the ring behaved as if they were uncoupled.  As the coupling
strength was increased the behavior of the ring became more coherent.
Since the discrete Laplacian acts to synchronize
neighboring systems in the ring, these results agree with our
expectations.  As was emphasized in the introduction, the onset of
the coherent behavior occurred at coupling strengths which seemed to be
independent of the number of oscillators in the ring. 
Some of the typical structures observed in these
numerical experiments are discussed below.

\subsection{Breathers}

If some of the systems in the ring are nearly stationary,
while others are undergoing oscillations we say that the oscillating systems
form a breather.  Since we are considering rings of
finite size, it is not expected that any single system in the ring is
stationary, unless the entire ring is stationary. 

\begin{figure}[th] 
\epsfig{figure=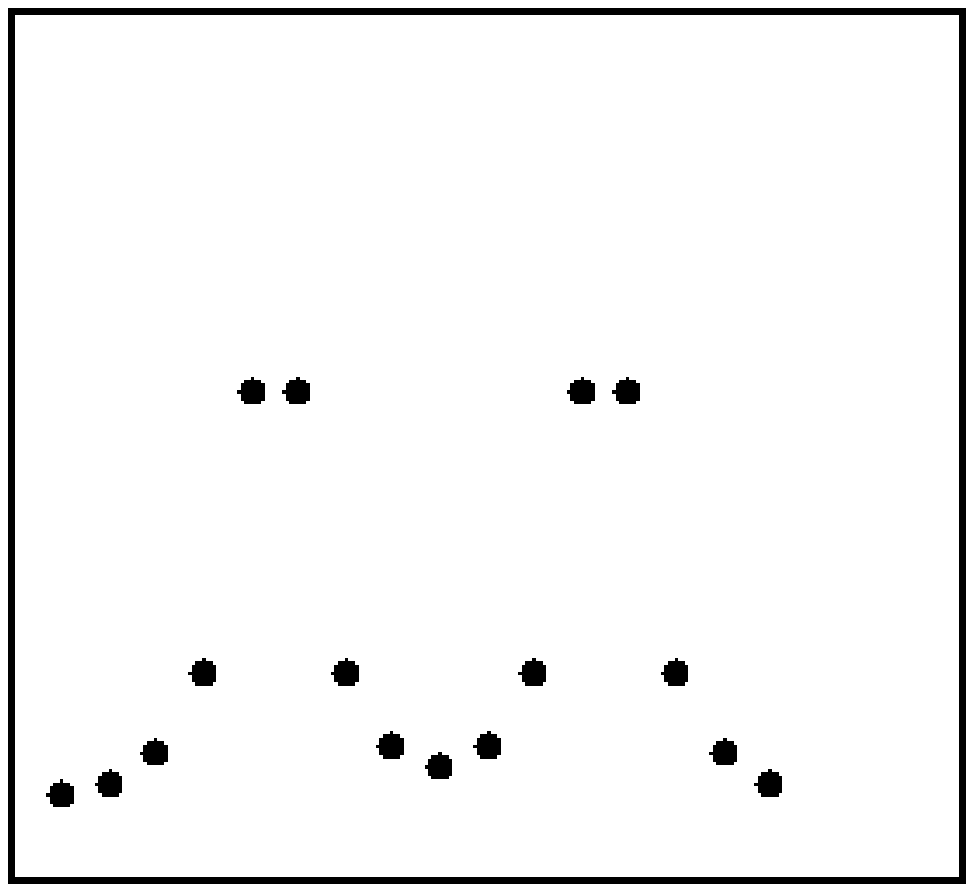, 
width=.32\linewidth, height = 40mm} \;
\epsfig{figure=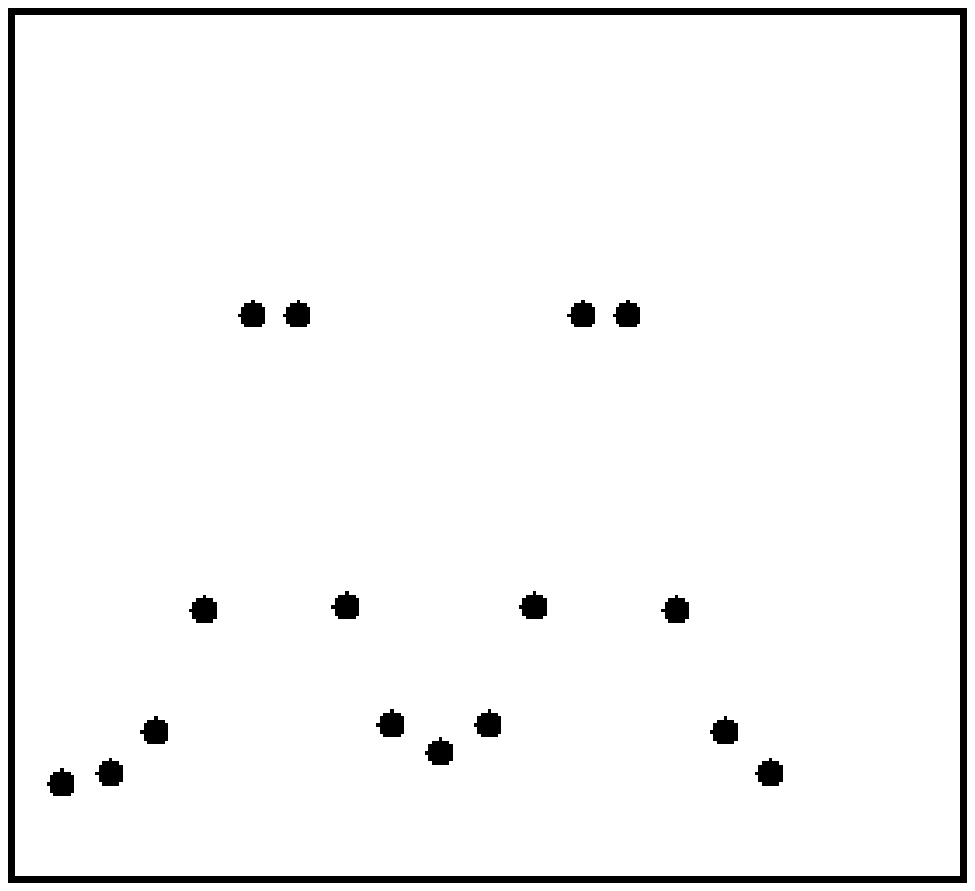, 
width=.32\linewidth, height = 40mm} \;
\epsfig{figure=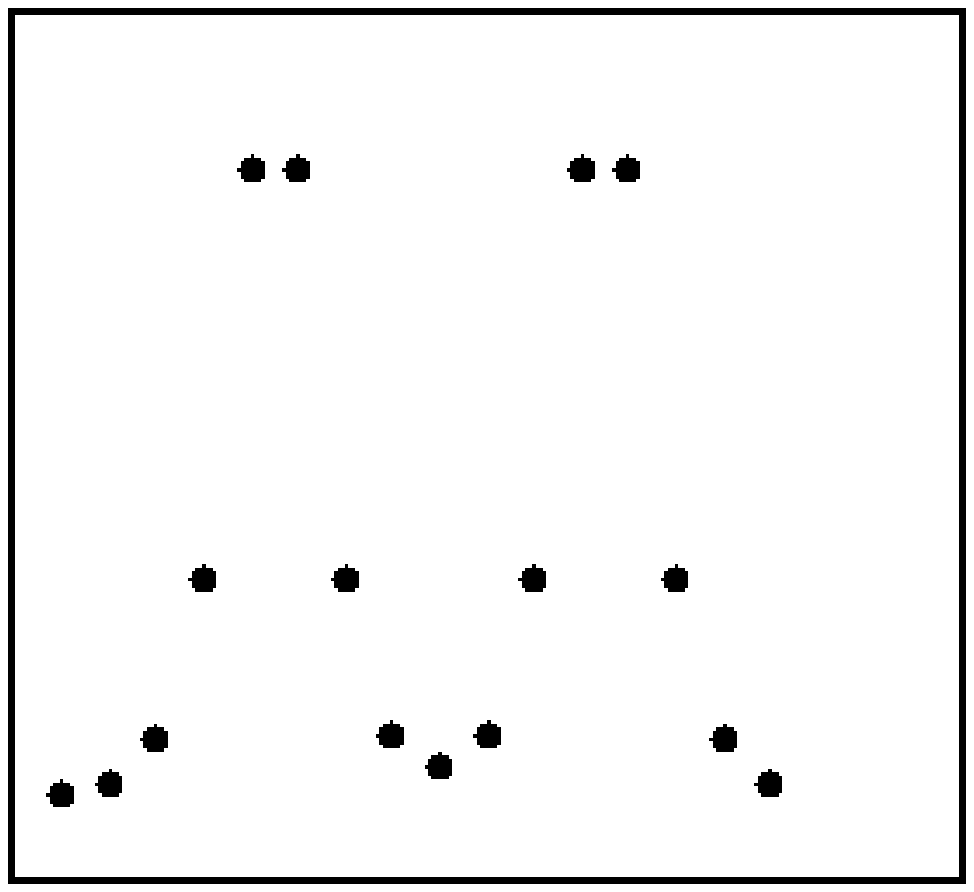, 
width=.32\linewidth, height = 40mm} 
\epsfig{figure=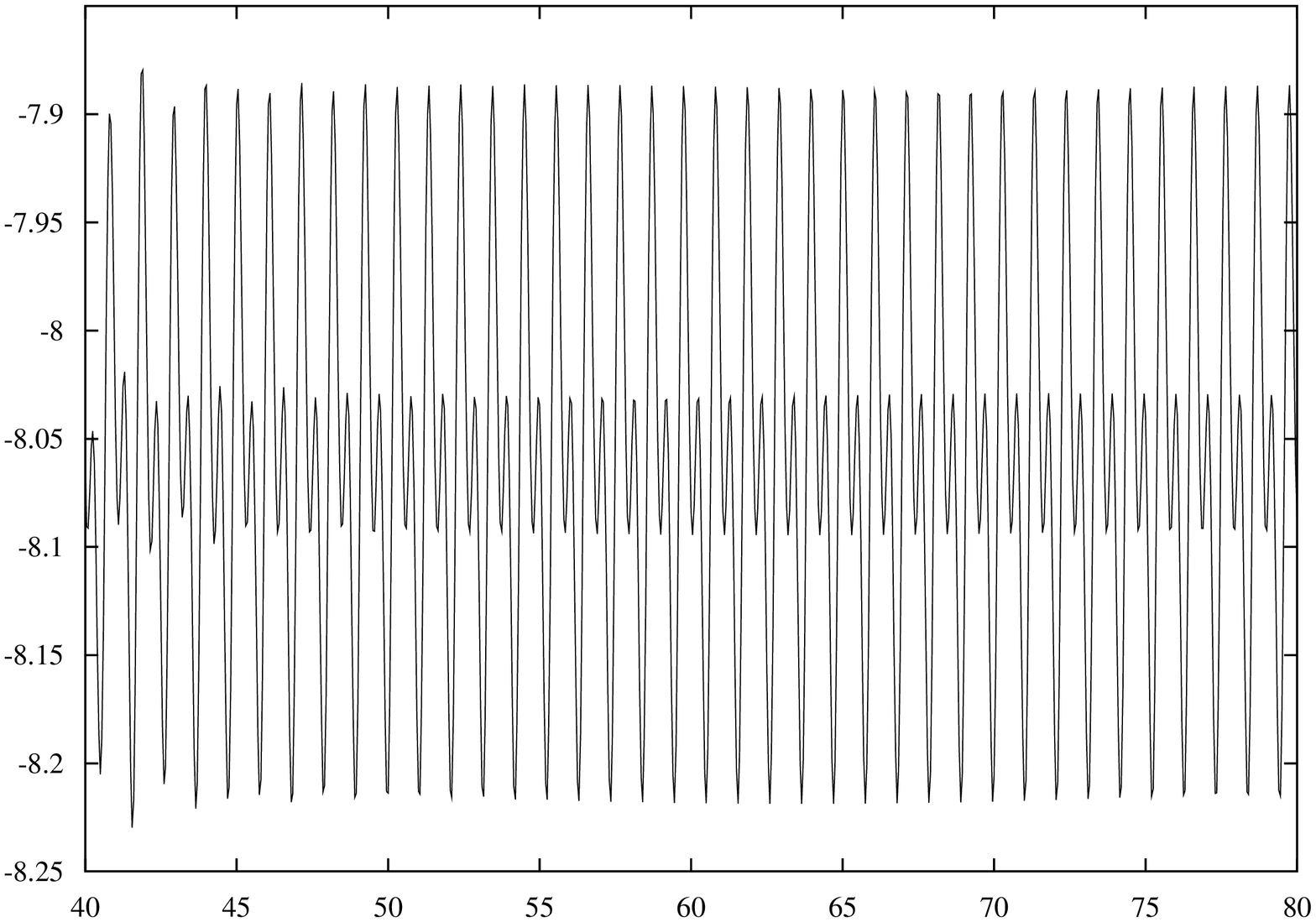,width=20mm, height = \linewidth, angle=-90}\\
\epsfig{figure=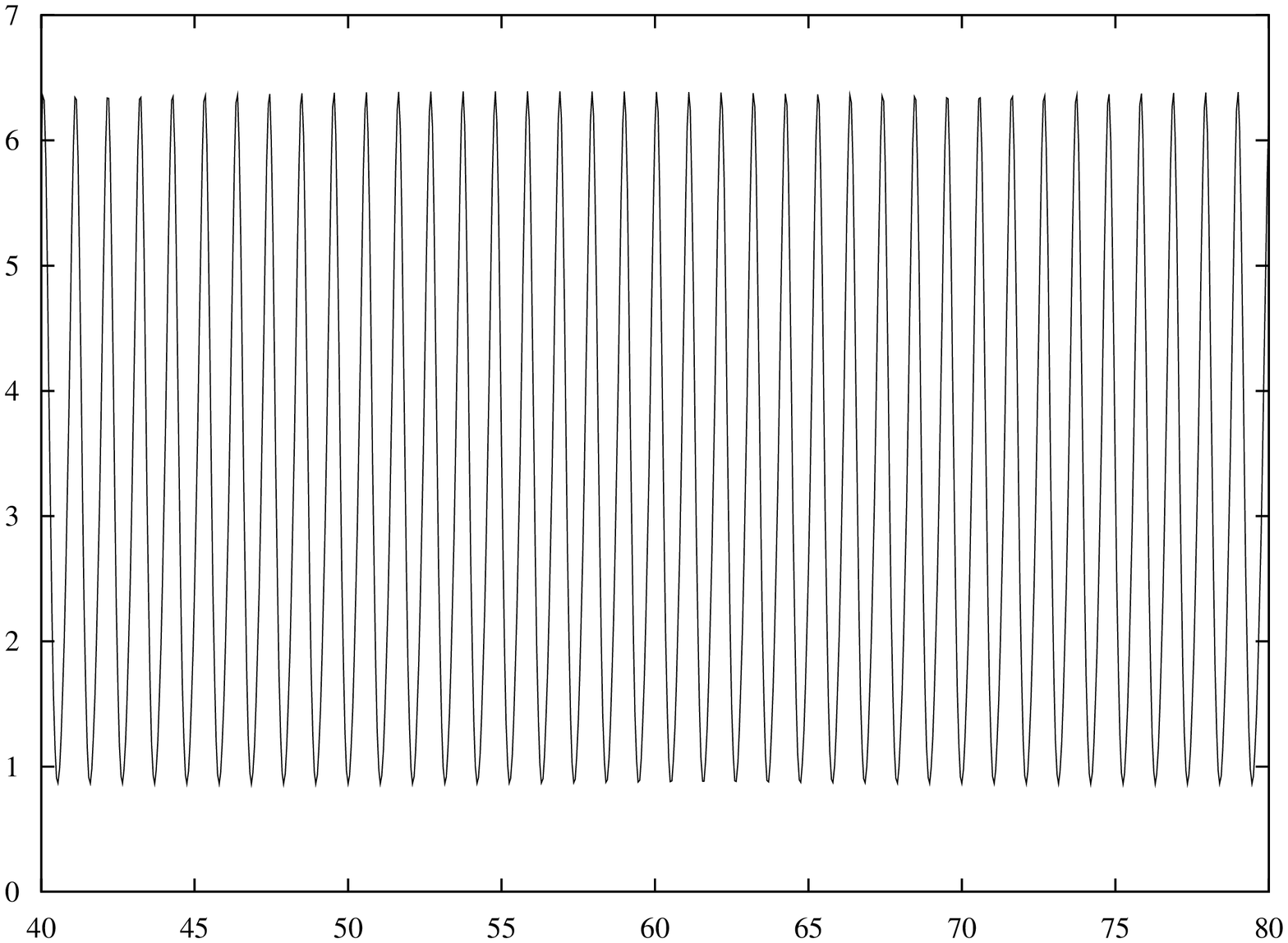,width=20mm, height = \linewidth, angle=-90}\\
\caption{A stable state of two breathers in a ring of 16
Lorenz systems coupled through the $x$ variable at $d_x = 15$.  
The first series of figures
shows a snapshot of the $x$ variables of the different systems at
different points of one oscillation.  In these figures, the position
of the oscillator in the ring is given on the $x$-axis, while $x_i(t)$
is given on the $y$-axis.  The timeseries of $x_1(t)$
and $x_5(t)$, which are nearly periodic, are given below.}\label{breath}
\vspace{5mm}
\end{figure}

An example of a system with two breathers is shown
in figure~\ref{breath}.  It is interesting to note that in these states
each of the individual systems undergoes oscillations that are nearly
periodic, and that there is a strong spatial dependence on the motion.
Loosely speaking the motion of each individual system is trapped in a region
close to one of the `lobes'' of the Lorenz attractor.  No stable states in
which all systems in the ring are trapped in the vicinity of only
\emph{one}  ``lobe'' has been observed.  
Similar breathers have been studied in coupled map lattices in
\cite{bunimovich}, while occurrences of nearly periodic behavior with
pronounced  spatial patterns in lattices of
chaotically bursting neurons have been observed numerically by
M.I. Rabinovich, et. al. \cite{rabinovich}.

Another interesting feature of breathers, which is also shared with
other types of solutions of this network, is that frequently
nonadjacent systems synchronize without
synchronizing with the oscillators that lie between them.
The synchrony is usually not exact, however this may be due to
numerical errors in the calculations.  Due to the many symmetries of
the system this sometimes seems to be due to the stability of an
invariant manifold corresponding to a partially synchronized state in the chain
as discussed \cite{josic}.  In other cases the state
does not appear to be symmetric, and it is unclear what the mechanism
behind this synchronization is. 

\begin{figure}[th]
\centering\epsfig{figure=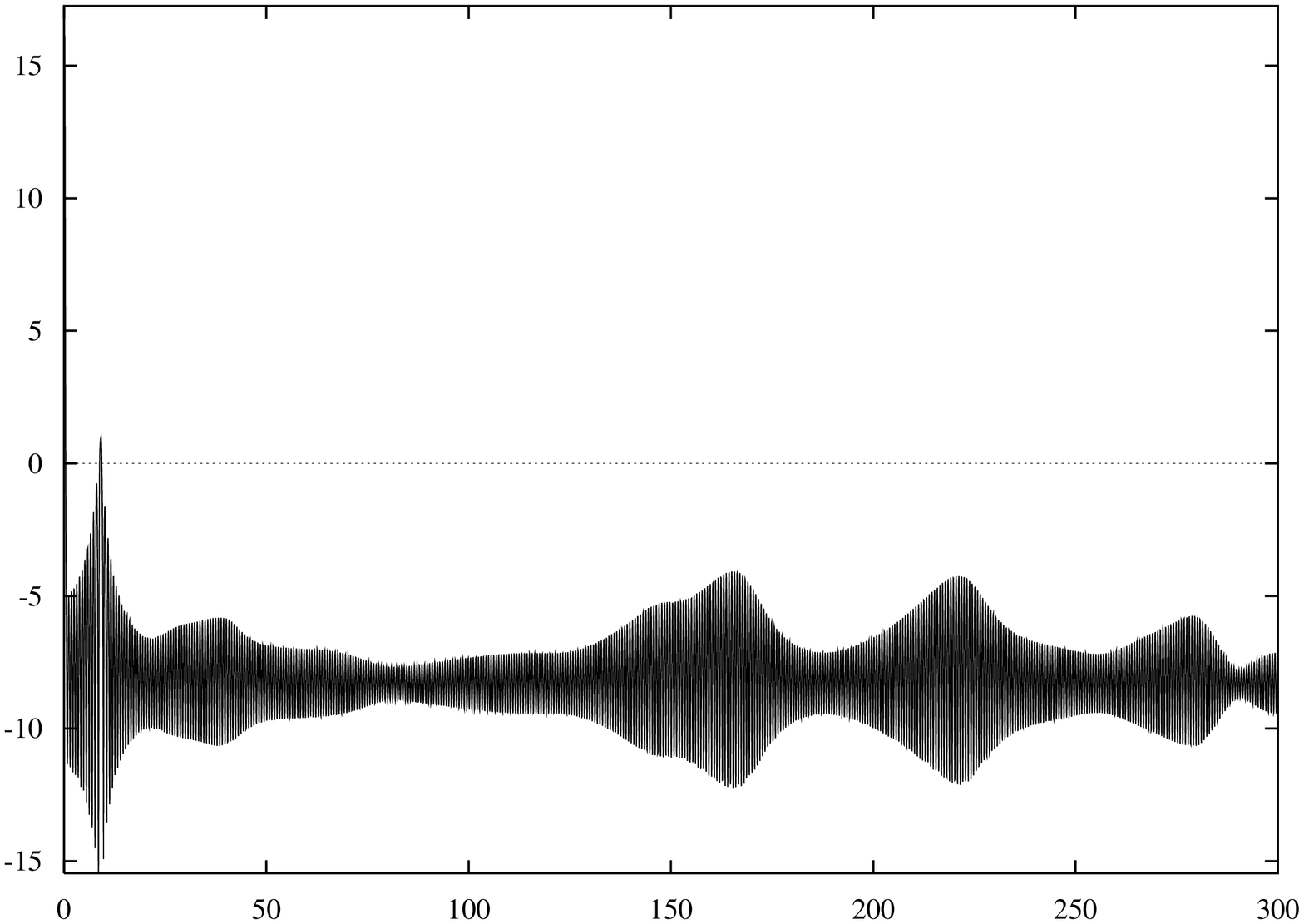, width=20mm, height = \linewidth, angle=-90}\\
\centering\epsfig{figure=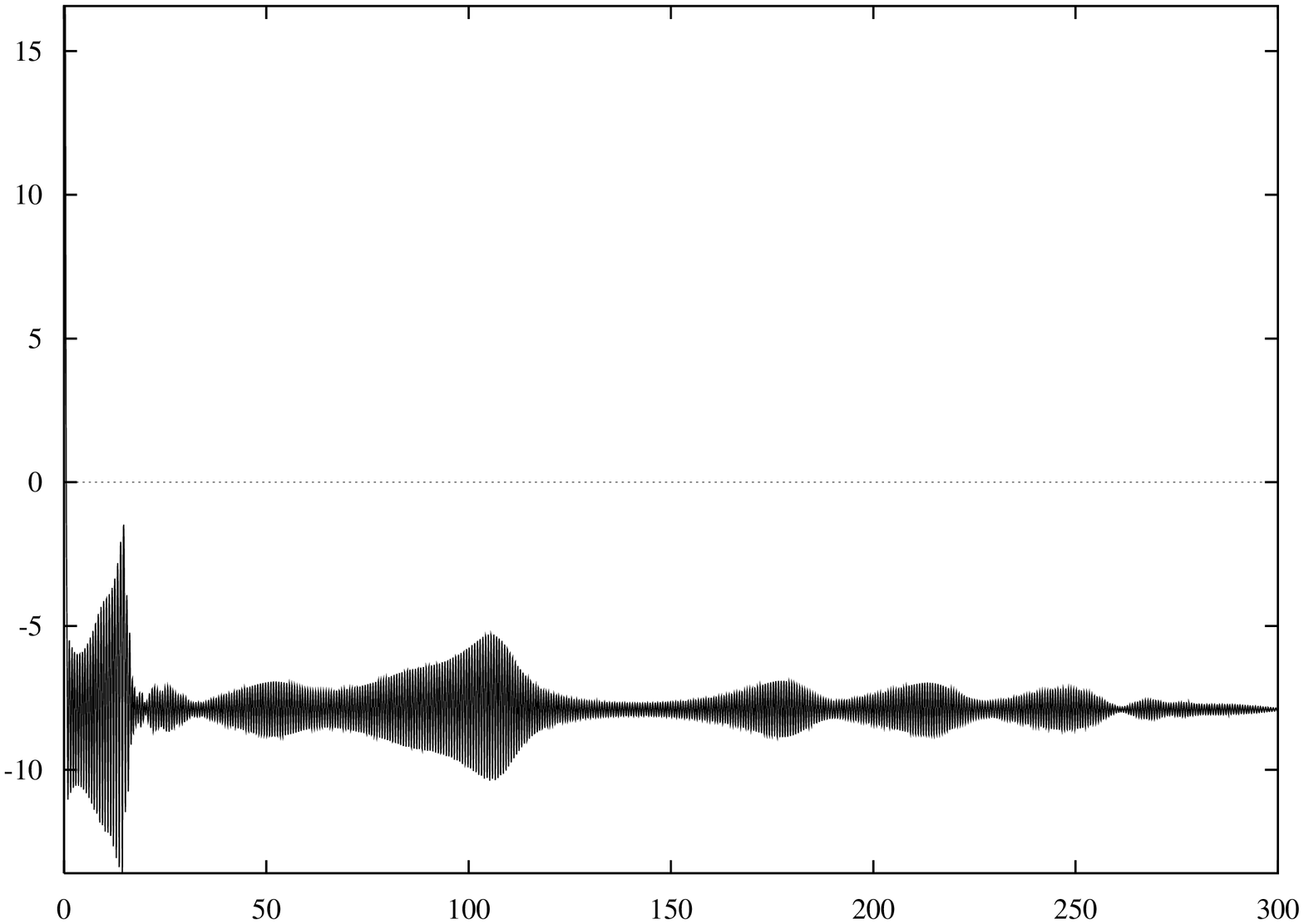,width=20mm, height =
\linewidth, angle=-90}\\
\centering\epsfig{figure=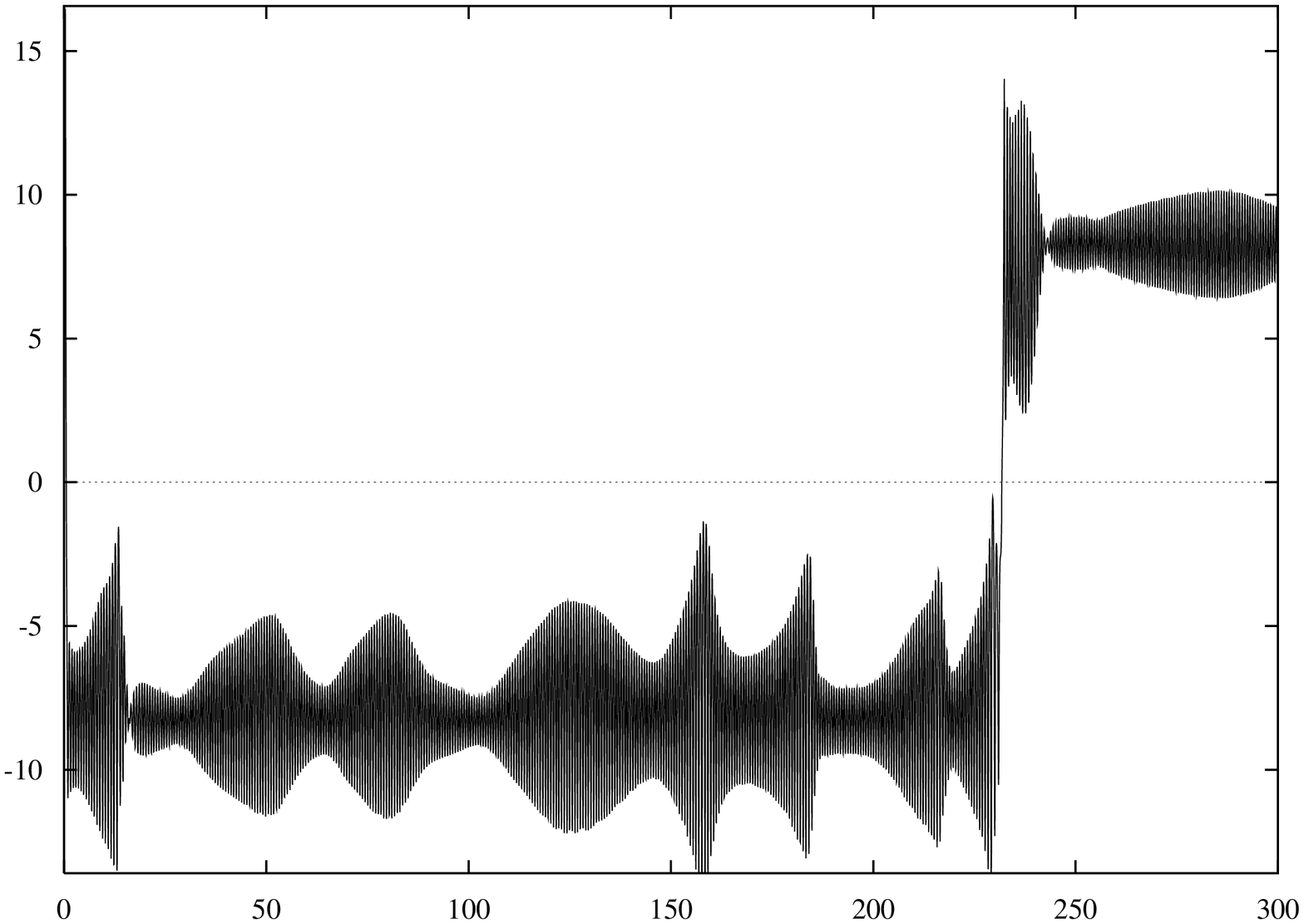,width=20mm, height =
\linewidth, angle=-90}\\
\centering\epsfig{figure=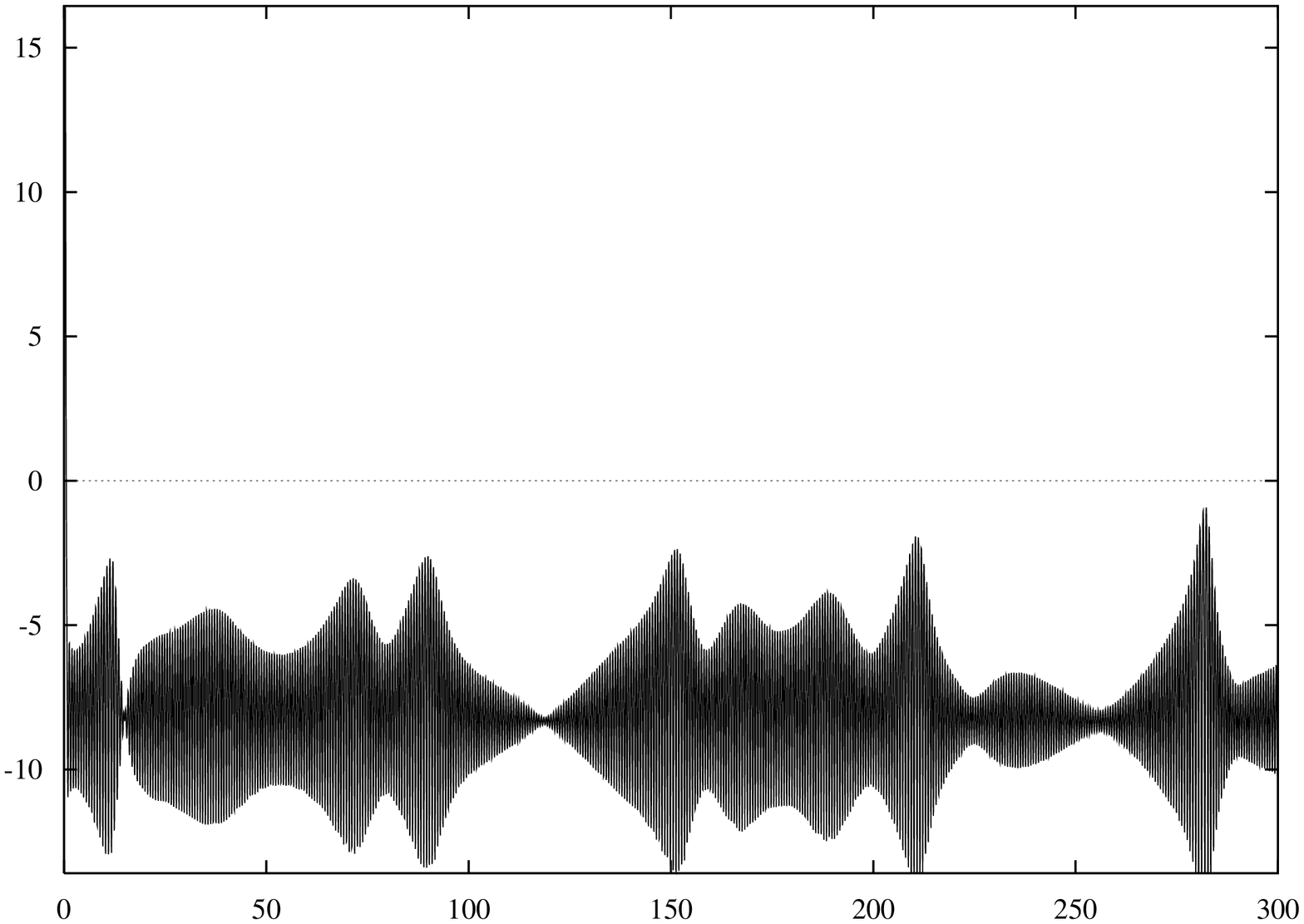,width=20mm, height =
\linewidth, angle=-90}\\
\caption{Typical timeseries of $x_{15}(t)$ at coupling strengths 
$d_x = 5, 15, 20, 25$ from top to bottom.  Many
features of the system remain similar at different coupling
values. The oscillations have a smooth envelope, a feature 
that is typical to solutions for coupling strengths at which the
network behaves coherently, but is not synchronized exactly.  
Even the transients that precede a stable fixed state have similar shapes.} \label{timese} 
\end{figure}

It is important to note that some of the qualitative features of the
dynamics remain unaffected by the size, or the coupling strength in the
ring.  Figure \ref{timese} shows typical timeseries of the $x$
variable at four different coupling values in a ring of 32
oscillators.  Note the smooth envelope and regularity of the
oscillations, as well as the agreement of the timescales among the
different timeseries.  These general features remain unchanged as the
number of systems in the ring is varied, as long as the coupling is
below the synchronization threshold and above the value necessary for
the appearance of coherent behavior.

\subsection{Stable Stationary States} 

\begin{figure}[th] 
\vspace{5mm}
\epsfig{figure=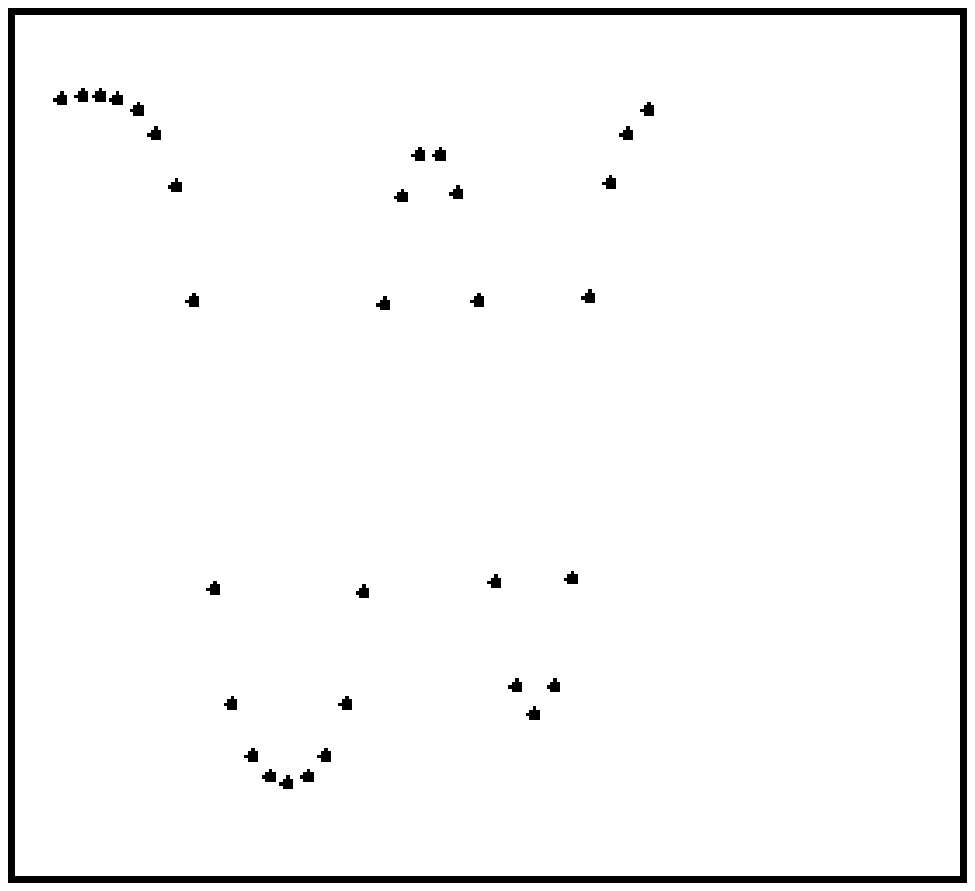, 
width=.32\linewidth, height = 40mm} \;
\epsfig{figure=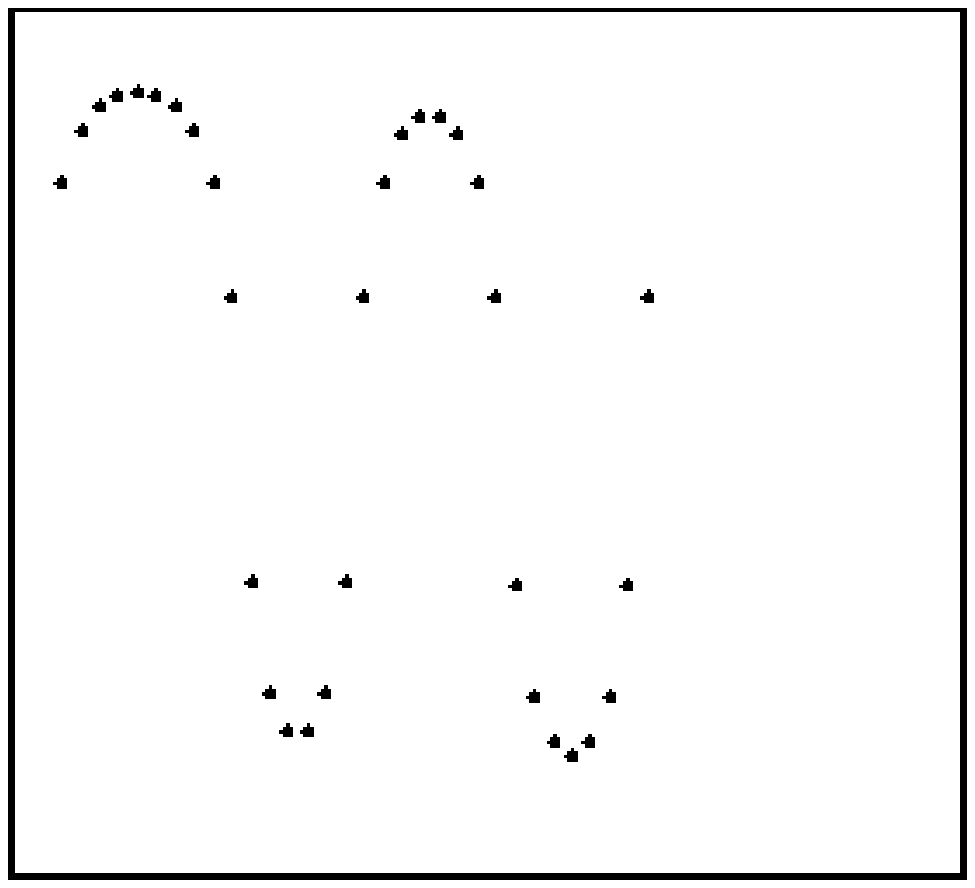, 
width=.32\linewidth, height = 40mm} \;
\epsfig{figure=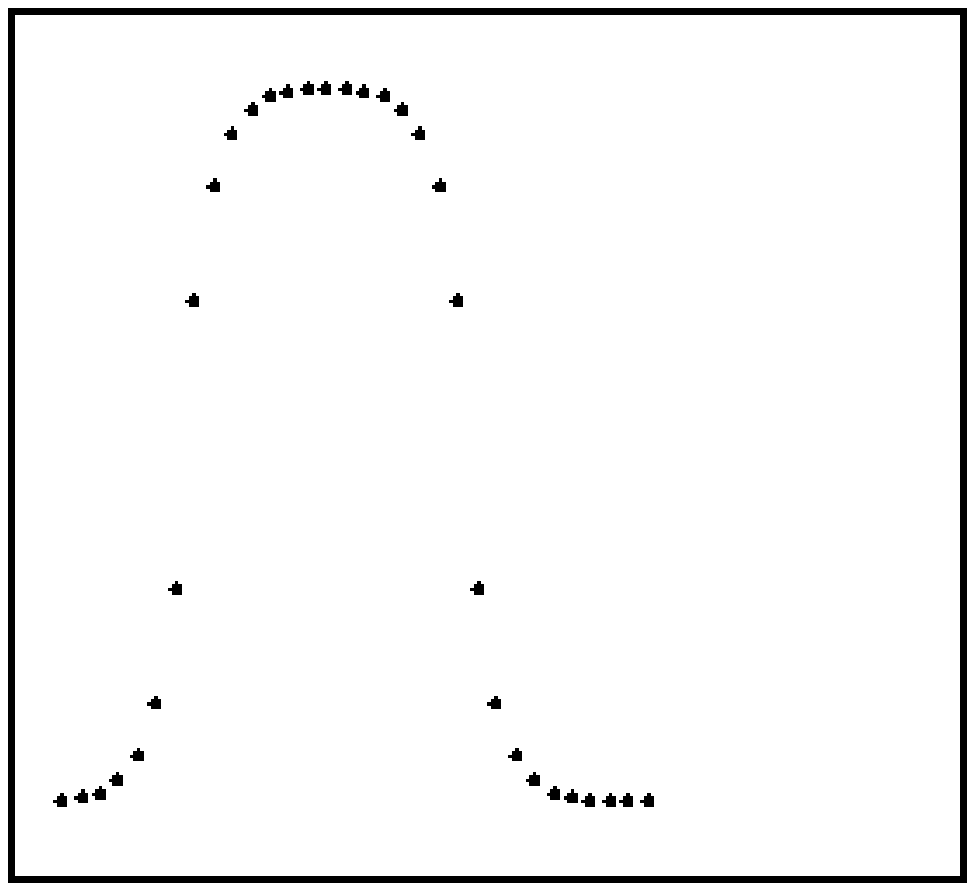, 
width=.32\linewidth, height = 40mm} 
\caption{The system is very sensitive to initial conditions.  The
eventual stable state in a chain of 32 oscillators with $d_x = 40$ is
shown.  Only one initial condition $x_0(0)$ is different for each
figure - it was incremented by 0.01.  Very different states are
reached.  This suggests that the basins of attraction are intertwined
in a complicated way.  The stable states in these figures are typical,
although fixed states with one ``hump'' are more common than states
with more ``humps''.}\label{sensitive}
\end{figure}

In numerical experiments with rings of 8 to 52 
systems stable steady states were observed for a variety of initial
conditions for $d_x > 10$.  
Typical stable fixed states are shown in figure \ref{sensitive}.   

For certain values of the coupling it appears that all
initial conditions lead to one of the stationary states of the
system.  The basins of attraction of the stationary states seem to be
intertwined in a complicated way in this case, as shown in figure
\ref{sensitive}.  The same is true for other stable states of the
system.  Small changes in the initial values of the parameters, and
changes in the integrating methods can result in very different
asymptotic behavior of the system.  This situation is similar to
that of attractors with riddled basins of attraction \cite{ashwin1},
and was also observed in numerical simulations of networks of
chaotically bursting neurons \cite{huerta}.

\subsection{Traveling Pulses}

\begin{figure}[th] 
\vspace{5mm}
\epsfig{figure=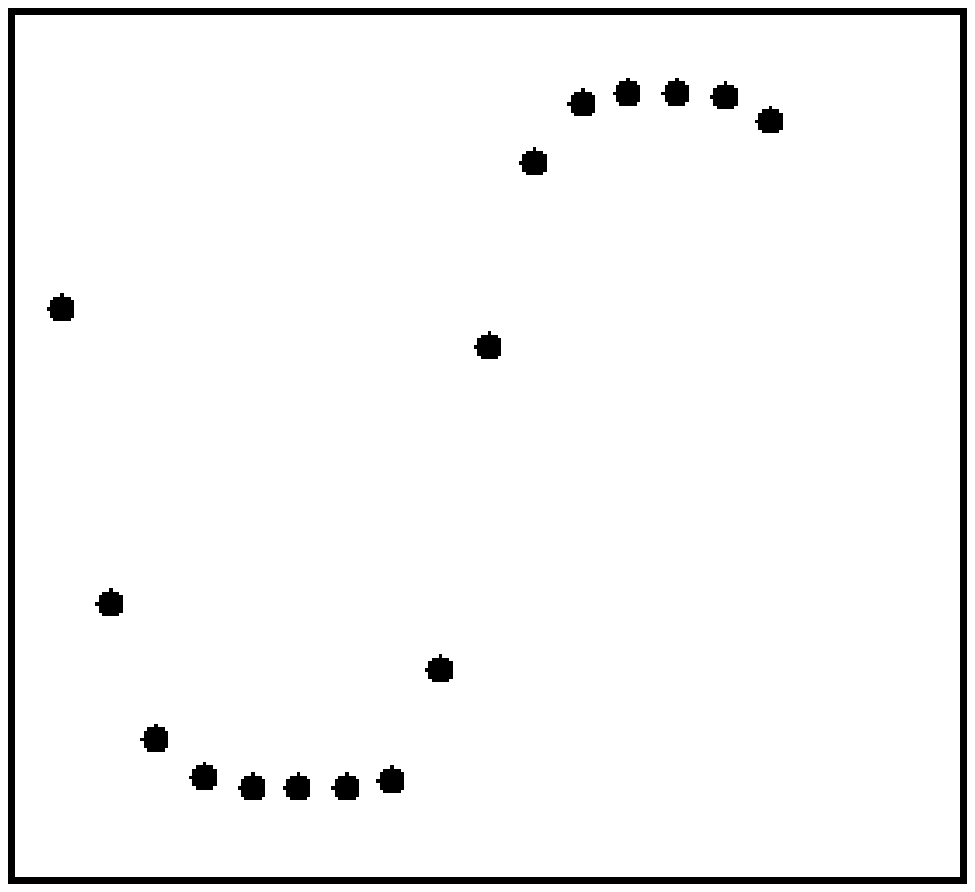, 
width=.23\linewidth, height = 30mm} \;
\epsfig{figure=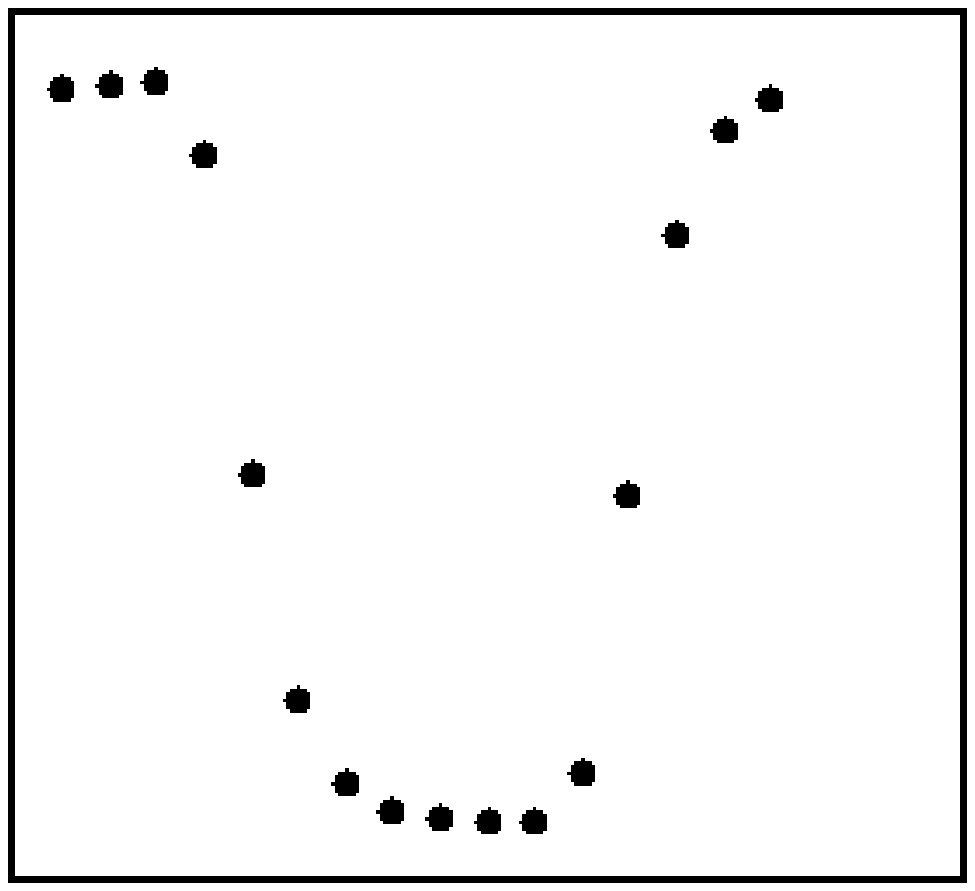, 
width=.23\linewidth, height = 30mm} \;
\epsfig{figure=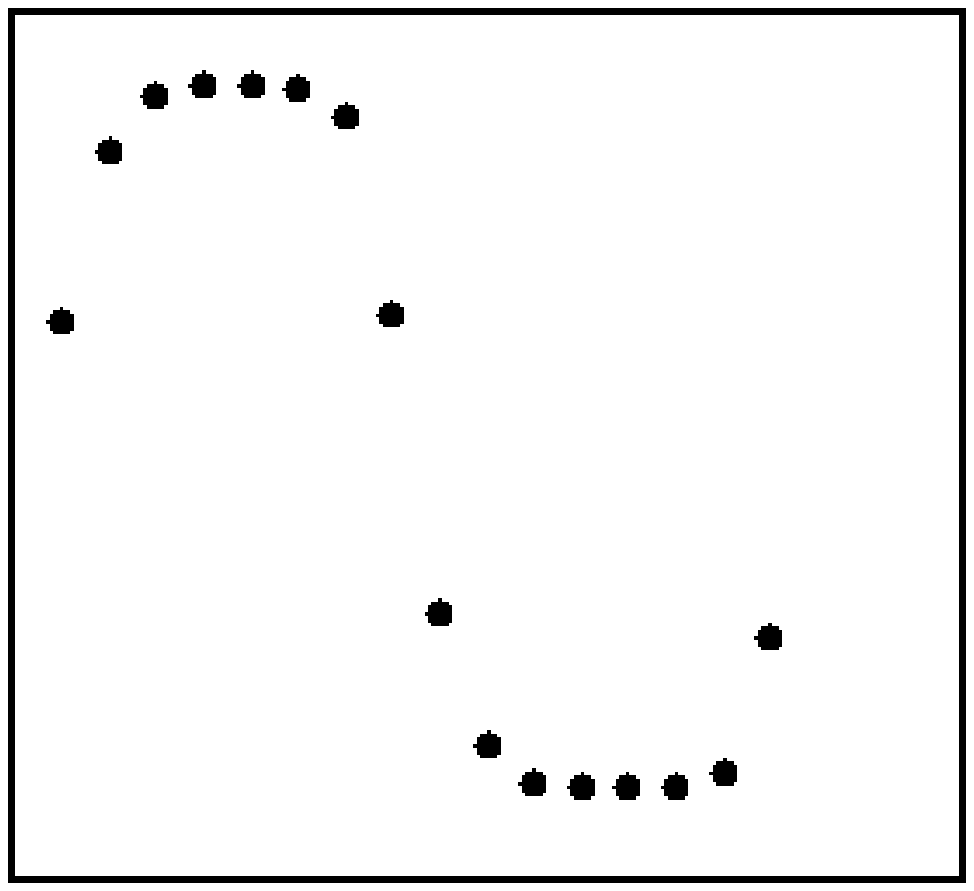, 
width=.23\linewidth, height = 30mm} \;
\epsfig{figure=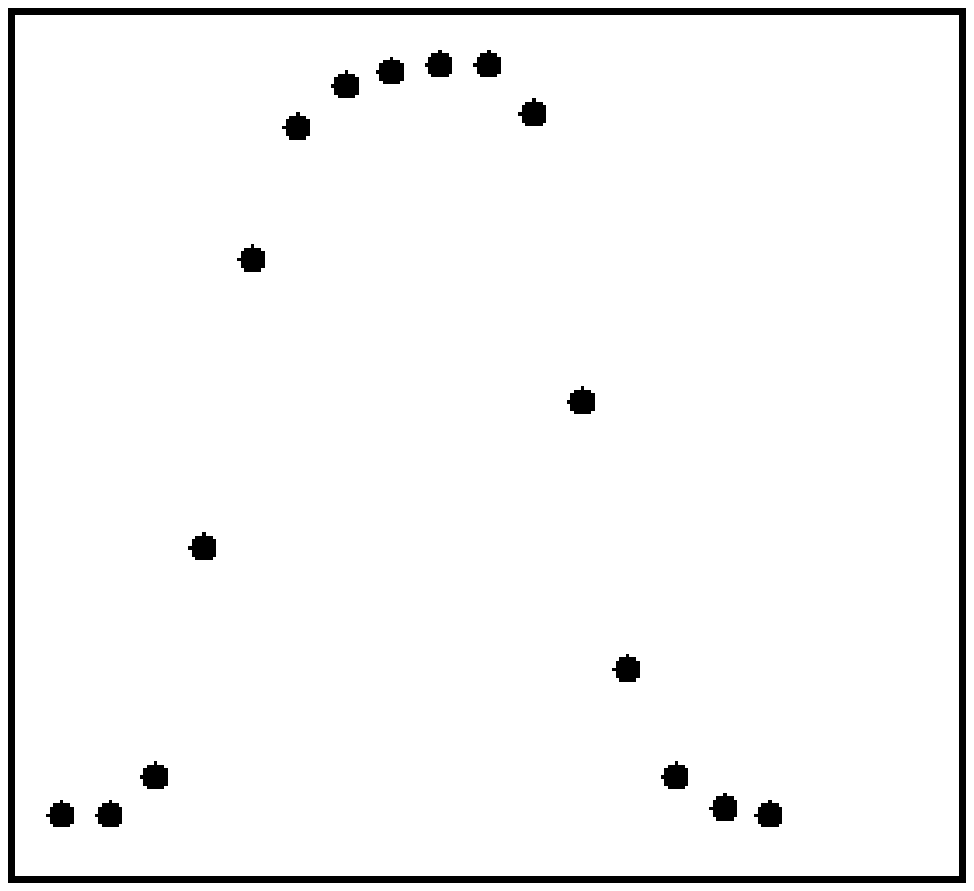, 
width=.23\linewidth, height = 30mm}
\caption{The wave when $d_x = d_y = 6$.  This seems to be a globally
stable solution of the equations.}\label{wave}
\end{figure}

Traveling pulses were observed in the ring only when $d_x \neq 0$ and 
$d_y \neq 0$.  A typical traveling pulse is shown in figure
\ref{wave}.  The pulse  oscillates as it
propagates along the chain, and is thus a periodic, rather than a
fixed state in a moving coordinate frame. This can be seen in the timeseries 
in figure \ref{wave}.

\section{A General Framework for the Study of Steady States}\label{steadystate}

System (\ref{coupledlorenz}) is a special case 
of a lattice discrete or continuous time dynamical system 
\begin{equation}\label{lattice}
(u_j)' = F(\{ u_j \}^s) 
\end{equation}
where $j \in \Bbb{Z}^D$, $u_j \in \Bbb{R}^p$ and $\{u_j\}^s =
\{u_i : |i-j| \leq s\}$.  In the following discussion we will
use the convention that in $u_i^k$ the subscript $i \in \Bbb{Z}^D$
denotes the position in the lattice, while the superscript $1 \leq k \leq p$ denotes
the component of the vector $u_i$.  Such systems have been 
studied by many authors \cite{AfChow}, \cite{BuSi}, \cite{Radu} as discrete
versions of partial differential equations of evolution type.

We will mainly be concerned with systems continuous in time  with $D=1$ and
$s=1$. These are simply chains of systems coupled to their nearest neighbors.
The special case of \emph{rings}, that is chains of finite size with periodic
boundary condition will be the focus of our attention.   
A state in a ring of
$n$ systems corresponds to a state of period $n$ in the spatial variable in 
an infinite chain. 

A steady state of a chain given in (\ref{lattice}) is determined by
\begin{eqnarray}
F(\{u_j\}^s) = F(u_{j-1}, u_j, u_{j+1}) = 0 \label{continuous}\\  
F(\{u_j\}^s) = F(u_{j-1}, u_j, u_{j+1}) = u_j\label{discrete}
\end{eqnarray}
in the continuous, respectively discrete case.  We will study systems
that satisfy the following condition:

{\bf Condition 1} If we write $F: \Bbb{R}^{3p} \rightarrow \Bbb{R}^p$
as $F(\chi, \eta, \zeta) = (F^1(\chi, \eta, \zeta),$ $  F^2(\chi, \eta, \zeta),$ $  \dots, $ $F^p(\chi, \eta, \zeta))$
then det $\!\left[ \frac{\partial F^i}{\partial \zeta^j}(\chi,\eta,\zeta) \right]_{i,j} \neq 0$.

By the Implicit Function Theorem in the case
of continuous time if det $\!\left[ \frac{\partial F^i}{\partial \zeta^j}(\chi,\eta,\zeta) \right]_{i,j}$ $ \neq 0$
for all values of $\chi,\eta,\zeta$ then either there exists a point $(a,b,c) \in \Bbb{R}^{3p}$ such that
$F(a,b,c) = 0$ and 
therefore a function
$G(\eta,\chi)$ such that $F(\chi,\eta,G(\eta,\chi))=0$, or $F(\chi,\eta,\zeta) = 0$ has no solutions and there are no
fixed points of (\ref{lattice}).  If such a function $G$ exists and we define 
$u_{j+1} = G(u_j, u_{j-1})$, $x_j = u_j$ and $y_j = u_{j-1}$ this leads to the following dynamical system
on $\Bbb{R}^{2p}$:
\begin{eqnarray}\label{fixed}
x_{j+1} = G(x_j , y_j) \\
y_{j+1} = x_j \nonumber 
\end{eqnarray}
The steady states of (\ref{lattice}) are given by the $x$-coordinates
of the orbits of (\ref{fixed}).  In general, the function $G$ may not
be unique, in which case more than one system of the form
(\ref{fixed}) is needed to determine all the fixed points of the chain.  
An equivalent argument holds in the case of
discrete time.

Since the function $G$ can assume any form, not much can be said about
such systems in general.  The subclass of chains of systems with the
following type of coupling is easier to analyze.
\begin{definition}
A nearest neighbor coupling of a chain of systems is said to 
be \emph{symmetric} if  $F(u_{j-1}, u_j, $ $u_{j+1})$ $=  F(u_{j+1}, u_j, u_{j-1})$.
\end{definition}
This simply means that a system in the chain is coupled to its left
and right neighbor in the same way. For instance
the couplings $\Delta u_j =  u_{j-1} - 2 u_j + u_{j+1}$ and
$\Psi u_j =  u_{j-1} u_{j+1} + u_j$ are of this type.
The stationary states of symmetrically coupled chains are related to
the following class of 
dynamical systems:
\begin{definition}\label{reversible}
Given a diffeomorphism
$\Phi:\Bbb{R}^{2p} \rightarrow \Bbb{R}^{2p}$ and an
involution $R: \Bbb{R}^{2p} \rightarrow \Bbb{R}^{2p}$ such that the
dimension of the fixed point set of $R$ is $p$ we say that
$\Phi$ is \emph{R-reversible} if
\begin{equation}
R \circ \Phi = \Phi^{-1} \circ R
\end{equation}
The dynamical system defined by 
$x_{i+1} = \Phi(x_i)$ is also said to be R-reversible.
\end{definition}
The relation between symmetrically coupled chains and reversible systems is given in the following
\begin{theorem} \label{rever}
The fixed states of a symmetrically coupled chain of systems
satisfying condition 1 correspond to the orbits of a dynamical system
of the form (\ref{fixed}) which 
is $R$-reversible and volume preserving.  
For such a system $R(x,y) = (y,x)$, with $x,y \in \mathbb{R}^p$.
\end{theorem}

\begin{figure}[th] 
\vspace{5mm}
\centering\epsfig{figure=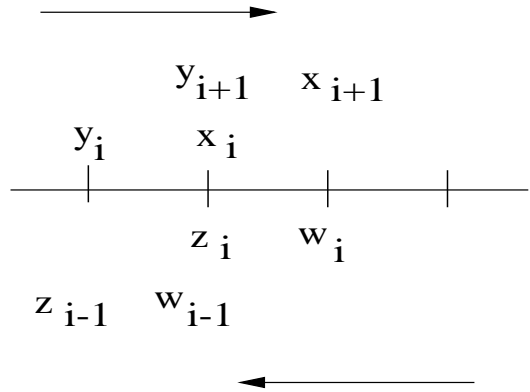, width=.5\linewidth,
height=50mm}
\caption {The action of the two systems in equation (\ref{dyn})}\label{action}
\end{figure}

{\bf Proof:} The arguments for the continuous and discrete case are
virtually identical, 
so only the first will be considered.  Equation (\ref{continuous}) together with the assumption that
the coupling is symmetric implies that
\begin{equation}
F(u_{j-1}, u_j, u_{j+1}) = F(u_{j+1}, u_j, u_{j-1})= 0
\end{equation}
Therefore this leads to the recursion equations
\begin{eqnarray}
u_{j+1} = G(u_j, u_{j-1}) & u_{j-1} = G(u_j, u_{j+1})
\end{eqnarray}
and the two dynamical systems
\begin{equation} \label{dyn}
\begin{aligned}
x_{j+1} &= G(x_j , y_j) \\
y_{j+1} &= x_j          
\end{aligned} \qquad \qquad 
\begin{aligned}
 z_{j-1} &= G(z_j , w_j)\\
 w_{j-1} &= z_j
\end{aligned}
\end{equation}
where $x_j = z_j = u_j$, $y_j = u_{j-1}$ and $w_j = u_{j+1}$.  The action of these
systems is shown schematically in figure \ref{action}.

Let $R(x,y) = (y,x)$ and $\Phi(x,y) = (G(x,y),x)$ so that the dynamical systems in (\ref{dyn}) is  generated
by $\Phi$. By the definition of this diffeomorphism
\begin{equation}
\begin{split}
R \circ \Phi \circ R \circ \Phi(x_i,y_i) & = R \circ \Phi (y_{i+1}, x_{i+1}) =
R \circ \Phi (z_i, w_i) \\
& = R (z_{i-1}, w_{i-1}) = R(y_i, x_i) = (x_i, y_i)
\end{split}
\end{equation}
Since $R$ is an involution with Fix$(R) = \{ (x,y) : x=y \}$ the diffeomorphism $\Phi$ and
the dynamical system it induces on the plane are $R$-reversible. 

To prove that the map $\Phi$ is volume  preserving notice that 
\begin{equation}
D\Phi = 
\begin{bmatrix}
D_1G & D_2G \\
I & 0
\end{bmatrix}
\end{equation}
By the definition of the function G we know that $F(\chi,\eta,G(\eta,\chi))=0$ so that
differentiating with respect to $\chi$ leads to 
\begin{equation}
D_1F + D_3F D_2G =0
\end{equation}
$D_3F$ is an invertible matrix by Condition 1, so that $D_2G = - D_1F
(D_3F)^{-1}$.  By assumption 
the system is symmetric so that $D_1F =D_3F$ at all points in $\Bbb{R}^{3p}$ and hence
$D_2G = -I$. Since
\begin{equation}
\begin{bmatrix}
0 & I \\
-I & 0
\end{bmatrix} 
\begin{bmatrix}
D_1G & -I \\
I & 0
\end{bmatrix} = 
\begin{bmatrix}
I & 0 \\
-D_1G & I
\end{bmatrix}.
\end{equation}
and
\begin{equation}
\det \begin{bmatrix} 0 & I \\
-I & 0
\end{bmatrix} =
\det \begin{bmatrix}
I & 0 \\
-D_1G & I
\end{bmatrix}=1
\end{equation}
it follows that $\det D\Phi =1$ and the diffeomorphism $\Phi$ is
volume preserving.
$\diamondsuit$

$R$-reversible systems have many properties that facilitate their study.
We will make use of several of these summarized in the
following proposition adopted from \cite{Dev}

\renewcommand{\theenumi}{\alph{enumi}}

\begin{proposition}\label{devaney}
Let $R$ and $\Phi$ be as in Definition (\ref{reversible}).
\begin{enumerate}
\item If $p \in$ Fix($R$) and $\Phi^k(p) \in$ Fix($R$), then $\Phi^{2k}(p) = p$.
Such periodic points will be referred to as \emph{symmetric periodic points}.
\item Let $p \in$ Fix($R$) be a fixed point of $\Phi$, then $R(W^u(p)) = W^s(p)$
and $R(W^s(p)) = W^u(p)$ so that if $q \in W^u(p) \cap$ Fix($R$) then $q$ is 
a homoclinic point.
\item If $p \in$ Fix($R$) is a fixed point of  $\Phi$ such that $W^u(p)$ intersects
Fix($R$) transversally at a point $q$, then there exist infinitely many symmetric
periodic points in any neighborhood of $p$.
\end{enumerate}
\end{proposition}

\section{Steady States In the Case of Discrete Laplacian Coupling} \label{section.steady}

For a chain of systems of the form $u' = f(u)$ coupled through a  discrete Laplacian $\Delta u_j =
(u_{j-1} - 2u_j + u_{j+1})$ the equation (\ref{lattice}) takes the form
\begin{equation}\label{lap}
(u_j)' = f(u_j) + d (u_{j-1} - 2u_j + u_{j+1})
\end{equation}
and the dynamical system (\ref{fixed}) takes the form
\begin{equation}\label{laplacefixed}
\begin{align}
x_{i+1} &= -\frac{1}{d} f(x_i) - y_i + (2 + \frac cd) x_i \\
y_{i+1} &= x_i \nonumber
\end{align}
\end{equation}
where $c=0$ in the case of continuous time and $c=1$ in the case of discrete time.
The following proposition follows immediately from these definitions
\begin{proposition}\label{lapfixed}
The fixed points of (\ref{laplacefixed}) are of the form $x_i = y_i = u_0$ where
$u_0$ is any solution of the equation $f(u) = 0$ in the continuous
time and $f(u)= u$ in the discrete time case.  Thus all fixed points of (\ref{laplacefixed}) 
lie on Fix$(R)= \{ (x,y) : x=y \}$ and are in 1-1 correspondence with the
steady states of an uncoupled system in the chain.
\end{proposition}

The fixed points of (\ref{fixed}) correspond to states of the 
chain that are constant in space and time and proposition~\ref{lapfixed}
shows that the only such states $\{u_j\}_{j \in \Bbb{Z}}$ in the case of a discrete Laplacian coupling
are given by $u_j = u_0$ for all $j \in \Bbb{Z}$ where $u_0$ is 
a fixed point of an uncoupled system of the chain $u' = f(u)$. Since the fixed points of (\ref{laplacefixed})
are $(u_0, u_0) \in$ Fix($R$)
this allows us to make use of proposition \ref{devaney}.c\footnote{The
chains under consideration are a special case of chains of the form
\begin{equation}\label{special}
(u_j)' = f(u_j) + g(u_{j-1}, u_j, u_{j+1}) \nonumber
\end{equation}
where $g(u_{j-1}, u_j, u_{j+1})$ is a symmetric coupling vanishing 
 on the linear submanifold
of $\Bbb{R}^{3p}$ defined by $u_{j-1} = u_j = u_{j+1} =0$.  Such
 systems can be analyzed using the approach of this section} . 

Moreover we can apply Theorem~\ref{rever} directly in this case to conclude that
the map defining the fixed points is $R$-reversible and volume preserving.

{\bf Remark:} It is interesting to note that if the fixed states
of a chain of one dimensional systems  are determined by a dynamical system 
of the form (\ref{laplacefixed})
on $\Bbb{R}^2$ and $(x_0, y_0)$ is a fixed point of (\ref{laplacefixed}) then
the discriminant of the characteristic polynomial of $D\Phi(x_0,y_0)$ is 
\begin{equation}
\sqrt{(-\frac{1}{d}f'(x_0)+ 2+c/d)^2 -4}.
\end{equation}
Since det~$D \Phi (x_0,y_0) =1$ we can conclude the following:
\begin{enumerate}
\item In the case of discrete time the eigenvalues of $D \Phi (x_0,y_0)$ are
on the unit circle when $f'(x_0) > 1$ and $d>(f'(x_0)-1)/4)$ or $f'(x_0) <1$ and $d<(f'(x_0)-1)/4)$.
\item In the case of continuous time, the eigenvalues of $D \Phi (x_0,y_0)$ are
on the unit circle for sufficiently large \emph{positive} $d$ when $f'(x_0) >0$ and
for sufficiently large \emph{negative} $d$ when $f'(x_0) <0$,  
\end{enumerate}
By a theorem of Birkhoff (see for example \cite{moser})
since $\Phi$ is area preserving, the dynamical system
(\ref{laplacefixed}) will generically have
infinitely many periodic orbits in any neighborhood of $(x_0, y_0)$.  Any such
periodic orbit will correspond to a fixed state of (\ref{lap}) which is periodic
in space, i.e. in the variable $j$. 

The condition  det $D \Phi (x_0,y_0) =1$ is not sufficient to guarantee that 
$D \Phi (x_0,y_0)$ will have eigenvalues on the unit circle if  (\ref{laplacefixed})
is not a dynamical system on the plane.  Additional information needs to be considered
to conclude the existence of such periodic orbits in this case.
\vspace{.5cm}

{\bf Example 1:} Consider a chain of tent maps coupled through a
discrete Laplacian described by the equation $u_j' = f(u_j) + \Delta u_j$ with
\begin{equation}
f(u_j) = \begin{cases}
         2 u_j & \text{if $u_j < 1/2$}, \\
         -2 u_j +2 & \text{if $u_j \geq 1/2$}.
         \end{cases}
\end{equation}
This is a discrete dynamical system with fixed states determined by
a homeomorphism of the plane given in (\ref{laplacefixed})
with $c=1$.  

The only fixed points of (\ref{laplacefixed}) are
(0,0) and (2/3, 2/3) by proposition~\ref{lapfixed}. Since in this case the dynamical system (\ref{laplacefixed}) is linear in each 
of the two halves of the plane $x < 1/2$ and $x \geq 1/2$ the analysis of the dynamics
around the fixed points is straightforward.  A direct calculation
shows that in the region $x < 1/2$ there is a family of invariant
ellipses whose major axis lies on the diagonal $D=\{(x,y) \in
\mathbb{R}^2: x=y \}$ and on which the action of (\ref{laplacefixed}) is a rotation.  As
$d$ is increased, these ellipses become more eccentric.

The fixed states of the chain corresponding to these orbits are not
dynamically stable. 
If $\hat{u}=\{\hat{u}_j\}_{j \in \Bbb{Z}}$ is a fixed state in the chain such that
$\hat{u}_j < 1/2$ for all $j$ then around this state the dynamics of the
chain are described by:
\begin{equation}
u_j' = 2u_j + d \Delta u_j
\end{equation}
The spectrum of the operator $2 + d \Delta$ is the interval
$[2-4d, 2]$ and so  $\hat{u}$ cannot be stable, and would not be
observed in numerical experiments.   

\begin{figure} 
\centering\epsfig{figure=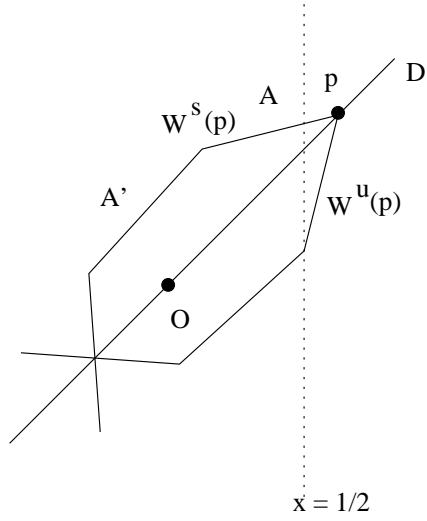, width=.4\linewidth}
\caption{The invariant circles around (0,0) give way to
complicated behavior close to 
(1/2, 1/2) due to the transversal intersection of $W^u(2/3, 2/3)$
and $W^s(2/3,2/3)$.}\label{ellman}
\end{figure}

Besides these invariant ellipses, this system will also typically have
infinitely many periodic orbits, and will exhibit complicated
behavior, as the following argument shows.  
A simple calculation shows that the point $p = (2/3, 2/3)$ is hyperbolic 
for all values $d>0$.  Since the dynamics
is piecewise linear the  stable and unstable manifold can be calculated 
explicitly.  $W^u(p)$ will consist of a ray contained
in the half plane $x>1/2$ and a second more complicated part which is
constructed as follows.  The first section of $W^u(p)$ is a line $A$,
as shown in figure \ref{ellman}.   
The part of $A$ contained in the half plane $x<1/2$ is
rotated along the invariant ellipses
around the origin to the line $A'$ under forward iteration.  
The subsequent images  of $A'$ are lines that are rotated further.
Therefore $f^{n_0}(A')$ intersects the diagonal $D$ for some $n_0$.
Whenever the angle of this intersection is not a right angle 
$W^u(p)$ will intersect $W^s(p)$ transversely since by 
proposition \ref{devaney}.b the stable manifold is the reflection
of the unstable manifold through the diagonal $W^u(p)$.  This shows
that complicated dynamics of (\ref{laplacefixed})
can be expected and implies the existence of infinitely many periodic
as well as spatially chaotic states in the chain of tent maps.
 
{\bf Example 2:}  Next we consider the chain of diffusively coupled 
Lorenz system (\ref{eq1}) with $d_y = d_z = 0$.  This system does not
satisfy condition 1, however after setting the right--hand side
of equations (\ref{eq1}) to 0 and
using the second and third equation in (\ref{eq1})
to eliminate the
variables $y_i$ and $z_i$  the following equation for a fixed
state is obtained 
\begin{align}
\sigma (\hat{x}_i - \frac{\beta r \hat{x}_i}{\beta + (\hat{x}_i)^2}) -
&  d  (\hat{x}_{i-1} 
- 2 \hat{x}_i + \hat{x}_{i+1})  = 0 \label{steady} \\ 
\hat{y}_i = & \frac{\beta r \hat{x}_i}{\beta + (\hat{x}_i)^2}  \\
\hat{z}_i = & \frac{r (\hat{x}_i)^2}{\beta + (\hat{x}_i)^2}.  
\end{align}
The function on the left hand side of equation (\ref{steady})
satisfies condition 1, and we can proceed as in the previous example.

Equation (\ref{steady}) defines the following dynamical system on
the plane  
\begin{equation} \label{lorenzfixed}
\begin{bmatrix}
x_{i+1} \\  y_{i+1}
\end{bmatrix} = F \left(
\begin{bmatrix}
x_i \\ y_i 
\end{bmatrix} \right) =
\begin{bmatrix}
\frac{\sigma}{d} (x_i -  \frac{\beta r x_i}{\beta + (x_i)^2}) + 2 x_i - y_i \\
x_i 
\end{bmatrix} 
\end{equation}
Orbits of period $n$ of this system are in 1--1 correspondence with
the  steady states in the ring of $n$ 
Lorenz systems. The two fixed points of (\ref{lorenzfixed}) are (0,0)
and $(\pm \sqrt{\beta (r-1)},\pm \sqrt{\beta (r-1)}) = (\pm
8.32\bar{6},\pm 8.32\bar{6})$ which both lie on
lie on Fix($R$).

The following definition from \cite{Dev} will be used in the remainder
of the argument.

\begin{definition}
A compact region $H$ in the plane is called \emph{overflowing} in the 
$y$-direction for a function $F$ if the image of any point $(x_0, y_0) \in$ int($H$)
lies strictly above the line $y=y_0$.  
\end{definition}

Notice that any point in the interior of $H$ either leaves $H$ or is asymptotic
to a periodic point on the boundary of $H$.  

\begin{theorem}
The map (\ref{lorenzfixed}) has a homoclinic point and infinitely many
periodic points for all $0<d \leq 20$.  
\end{theorem}

\begin{figure} 
\centering\epsfig{figure=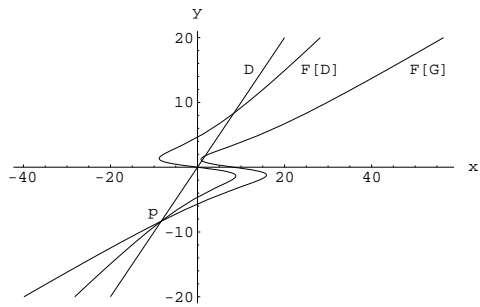, width=.45\linewidth}
\caption{The region $H$ is bounded by the two
cubics $F(D)$ and $F(G)$ and $D$.  The figure on the left W
corresponds to the case $d=20$.  For $d > 20+\epsilon$ these
curves do not bound a compact region. } \label{wedge2}
\end{figure}

{\bf Proof:}  Let $p$ denote the fixed point $(-\sqrt{ \beta (r-1)},$ $-\sqrt{ \beta (r-1)})$ 
of (\ref{lorenzfixed}), $G$ the ray $y = -\sqrt{ \beta (r-1)}$ with $x > -\sqrt{ \beta (r-1)}$,
and $D = \{(x,y): x=y\}$ the diagonal in $\Bbb{R}^2$.  The image of the
triangular region $W$ bounded by $D$ and $G$ lies between the 
two cubics
\begin{eqnarray}
F(D) = \{(x,y): x = \frac{\sigma}{d} (y - \frac{\beta r y}{\beta + y^2}) + y\} \\
F(G) = \{(x,y): x = \frac{\sigma}{d} (y - \frac{\beta r y}{\beta + y^2}) + 2 y  + \sqrt{ \beta (r-1)} \}
\end{eqnarray}
Consider the region $F(W) \cap W = H$ depicted in figure \ref{wedge2}.
Let $(x_0, y_0) \in$ int($H$)
and consider the vertical line segment $l$ passing through $(x_0, y_0)$ and connecting $G$
and $D$. The image of $l$ is a horizontal line segment connecting $F(G)$ and
$F(D)$ contained in the line $y = x_0$.  Any point of this line segment 
lies above the line $y = y_0$.  
Therefore $H$ is overflowing in the $y$ direction.

\begin{figure} 
\centering\epsfig{figure=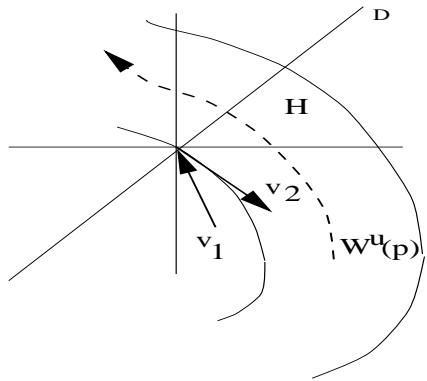, width=.4\linewidth, height = 50mm}\\
\caption{If $W^u(p)$ coincides with the stable manifold of the
origin then $v_1$ must be tangent to $W^u(p)$ at the origin.
Since $v_1$ points outside of $H$ this is impossible.} \label{vectors}
\end{figure}

Since int($F(H) -$ int($H$)) and int($H$) lie on opposite sides of $D$, any
point in $H$ must have an iterate which either lies on $D$ or crosses
$D$. A direct calculation shows that one branch of $W^u(p)$ enters
$H$. Since $H$ is overflowing in the $y$ direction, $W^u(p)$ must either
cross $D$ or be asymptotic to the fixed point $(0,0)$ creating
a saddle-saddle connection.  The second
possibility can be excluded as follows:

The eigenvector of the linearization of F around (0,0) corresponding to the stable direction
is $\mathbf{v}_1= (1, -\frac{d}{d-130+ 2 \sqrt{65^2 -d}})$.  The vector $\mathbf{v}_2=(1, \frac{d-260}{d})$ is tangent
to $F(D)$ at (0,0).  A direct computation shows that for $0<d<20$ the vector $\mathbf{v}_1$ points to the left of
$\mathbf{v}_2$ and so the tangent to $W^s(0,0)$ at (0,0) does not
point into $H$.  This situation is depicted schematically in figure
\ref{vectors} and excludes the
possibility that $W^u(p)$ is asymptotic to (0,0).

By proposition~\ref{devaney}.b the fact that $W^u(p)$ meets $D$ implies the existence of a homoclinic
point. 

By proposition \ref{devaney}.c in the case of $R$-reversible
systems an infinite number of periodic points exists whenever $W^u(p)$ crosses
$D$ transversely at a homoclinic point. If this point of intersection
is not transverse then a transverse intersection of $W^u(p)$
and $D$ can be produced nearby by an argument given in
\cite[p. 261]{Dev}. $\diamondsuit$

Numerical investigations suggest that the theorem remains true for
arbitrarily large values of d.

\begin{figure} 
\vspace{-1.5cm}
\centering\epsfig{figure=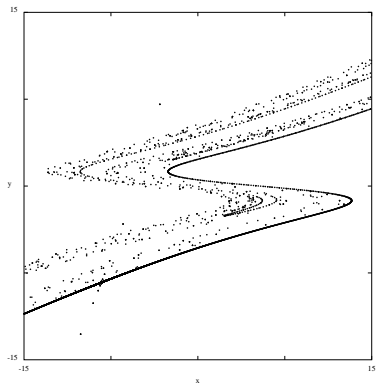, width=.45\linewidth}
\vspace{-1cm}
\caption{A numerical approximation of the manifold $W^u(p)$ is shown on the right.
The complex structure of the manifold suggests complicated
behavior on some invariant set of points. } \label{stablemanofp}
\end{figure}

This theorem shows that rings of Lorenz systems coupled through the
first variable can be expected to have many periodic stationary
states. Numerical computations of  $W^u(p)$ suggest that even more
complicated stationary states can be expected (see figure
\ref{stablemanofp}).  However
it is unclear whether any of them are stable.   
Numerical investigations show that not only are they stable, but
their basins of attraction occupy a large portion of phase space. 
We next address the problem of stability of these fixed states. 

\section{Stability of Steady States in a Ring} \label{stabilityofsteady}
 
The study of stable fixed states in a chain of oscillators is not new.
They are described in \cite{erm}, \cite{aron} as instances of oscillator
death in a chain of neural oscillators, in \cite{nekor} as synchronous
states of phase lock loops, and  conditions for the stability
of complex stationary states in general lattices are given in 
\cite{AfrChow2}.  The situation presented here is different in
that an explicit periodic stationary state is analyzed in a case where
the methods from the theory of parabolic partial differential
equations used in  \cite{erm} are not applicable, and the conditions
proposed in \cite{AfrChow2} cannot be verified.  In addition, since
the coupling strength is neither very large nor very small, there is no
obvious perturbative approach to the problem. 

Even if some of the fixed states in a ring of $n$ systems are 
stable it can not be concluded that stable equilibria exist in
arbitrarily long rings of such systems. If a stable fixed
state is viewed as a special case of synchronization, one might 
guess from results in Afraimovich and Lin \cite{lin} that the
coupling necessary to have stable steady states in longer rings
increases as a power of the size of the ring.  It has already been
argued that this is not realistic from a physical viewpoint, and we
will show below that it is not the case. 

In the following sections we will concentrate on the specific example of
a ring of Lorenz systems coupled in the $x$ variable discussed in example
2 of the last section.  However the ideas presented can be applied
to any chain of symmetrically coupled systems for which condition 1 holds.
 
The linearization of each of the Lorenz equations
 around a point $\mathbf{\hat{v}_i} = (\hat{x}_i, \hat{y}_i, \hat{z}_i )$ when
$d = 0$ is 
\begin{equation}\label{linlorenz}
D_{\mathbf{\hat{v}_i}}f = \begin{bmatrix}
-\sigma & \sigma & 0 \\
r - \hat{z}_i & -1 & - \hat{x}_i \\
\hat{y}_i & \hat{x}_i & - b 
\end{bmatrix} 
\end{equation}
and hence linearizing the entire ring around the steady state 
$\hat{\mathbf{v}} = ( {\mathbf{\hat{v}_i}} )_{i=1}^n = (\hat{x}_1,
\hat{y}_1, \hat{z}_1, \dots, $ $\hat{x}_n, $ $\hat{y}_n, $ $\hat{z}_n
)$ leads to the $3n \times 3n$ matrix
\begin{equation}
D_{\mathbf{\hat{v}}}F = \begin{bmatrix}
D_{\mathbf{\hat{v}_1}}f - 2\Gamma & \Gamma & 0 & \dots & 0 & \Gamma \\
\Gamma  & D_{\mathbf{\hat{v}_2}}f - 2\Gamma& \Gamma & \dots & 0 & 0 \\
\hdotsfor[3]{6}                                   \\
\Gamma   & 0 & 0 & \dots &  \Gamma & D_{\mathbf{\hat{v}_n}}f -2\Gamma 
\end{bmatrix}
\end{equation}
where the matrix $\Gamma$ is defined as:
\begin{equation}
\Gamma = \begin{bmatrix}
d & 0 & 0 \\
0 & 0 & 0 \\
0 & 0 & 0 
\end{bmatrix}
\end{equation}

\begin{figure} 
\centering\epsfig{figure=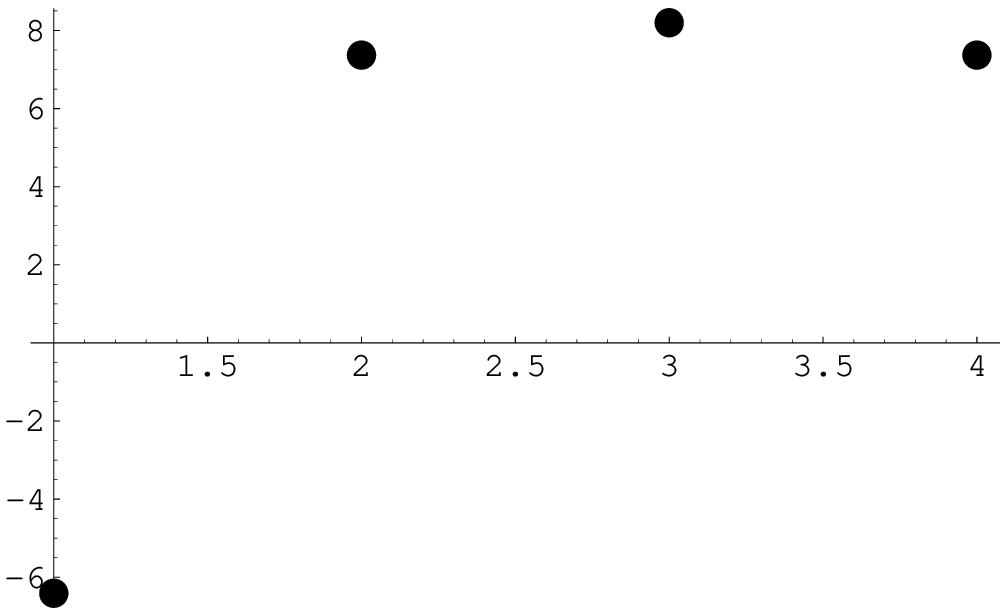, width=.45\linewidth} \qquad
\centering\epsfig{figure=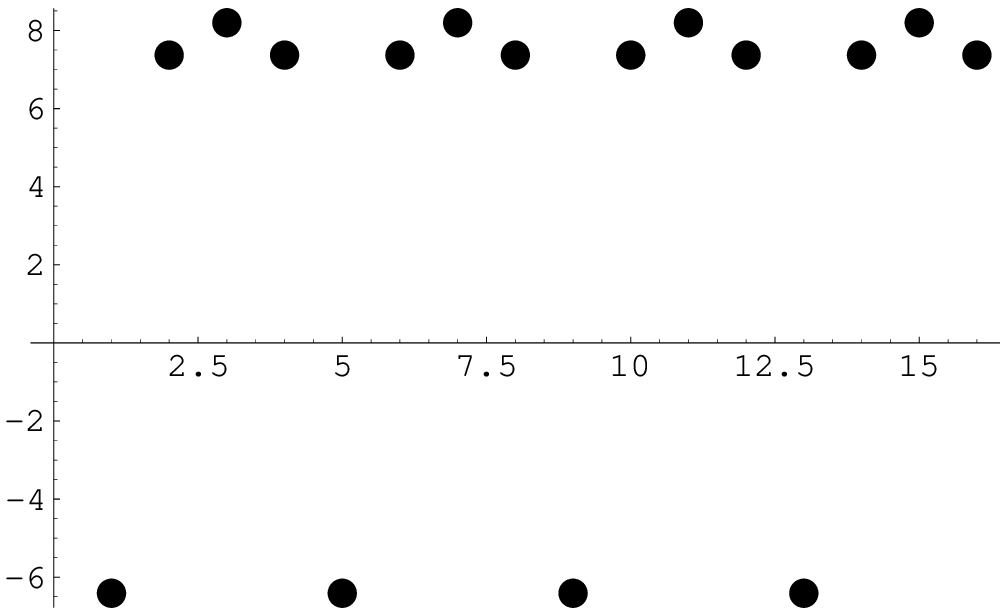, width=.45\linewidth}
\caption{The fixed state that will be used as an example, and its 4-multiple.}\label{kmultiples}
\end{figure}

It is in general not possible to compute the eigenvalues of this
matrix analytically. 
As mentioned above, we are specifically interested in coherent states
in networks with arbitrarily many oscillators.  Given a periodic
stationary state in a chain of $n$ oscillators, an easy way to obtain
a periodic stationary state in arbitrarily long chains is to repeat
this state.  
In particular, if $\mathbf{\hat{v}}$
is a fixed state in a ring of $n$ oscillators 
then repeating these values $K$ times to obtain 
$\mathbf{\hat{v}}^{(K)} = \{\mathbf{\hat{v}}, \mathbf{\hat{v}}, \dots, \mathbf{\hat{v}} \}$ produces a steady
state in a chain of $Kn$ oscillators.  We refer to such a state
as the $K$-{\em multiple} of the steady state $\mathbf{\hat{v}}$. An
example of a stationary state and its 4-multiple is given in figure
\ref{kmultiples}.
 We next
derive conditions under which all $K$-multiples of a stable steady
state are themselves stable and demonstrate that these conditions hold in a 
particular case.

The  linearization 
around the steady state $\mathbf{\hat{v}}^{(K)}$ leads to the following stability
matrix:
\begin{equation*}
D_{\mathbf{\hat{v}}^{(K)}}F = \begin{bmatrix}
M & E_d & 0 & \dots & 0 & E_u \\
E_u & M & E_d & \dots & 0 & 0 \\
\hdotsfor[3]{6}                                   \\
E_d & 0 & 0 & \dots & E_u & M 
\end{bmatrix}
\end{equation*}
where
\begin{equation*}
M  = \begin{bmatrix}
D_{\mathbf{\hat{v}_1}}f - 2\Gamma & \Gamma  & 0 & \dots & 0 & 0 \\
\Gamma  & D_{\mathbf{\hat{v}_2}}f - 2\Gamma & \Gamma & \dots & 0 & 0 \\
\hdotsfor[3]{6}                                   \\
0   & 0 & 0 & \dots &  \Gamma & D_{\mathbf{\hat{v}_n}}f -2\Gamma
\end{bmatrix}
\end{equation*}
and $E$ is an $3n \times 3n$ matrix of the form 
\begin{align*}
E_u = \begin{bmatrix}
0 & \dots & \Gamma \\
\hdotsfor[3]{3} \\
0 & \dots & 0 
\end{bmatrix} \qquad  &
E_d= \begin{bmatrix}
0 & \dots & 0\\
\hdotsfor[3]{3} \\
\Gamma & \dots & 0 
\end{bmatrix}
\end{align*}

In general it is impossible to compute the spectrum of this $3Kn
\times 3Kn$ matrix.  However in the present instance
Bloch wave theory,
provides a way of expressing the eigenfunctions of a larger system
with a periodic dependency as the eigenfunctions 
of a smaller system.  As usual, this reduction is accompanied by the
introduction of an additional parameter into the equations.  
Here the Bloch wave  approach will be used to reduce the problem of computing the 
eigenvalues of the $3Kn \times 3Kn$ matrix $D_{\mathbf{\hat{v}}^{(K)}}F$ for arbitrary $K$ to the computation of 
the eigenvalues of an $3n \times 3n$ matrix dependent on a parameter $t$.

Let $\mathbf{e} = (\Psi(j), \Phi(j), \eta(j) )_{j = 1}^{Kn} = (\Psi(1), \Phi(1), \eta(1), \dots, \Psi(Kn), \Phi(Kn), \eta(Kn)) $
 where
\begin{eqnarray}
\Psi(j) = \exp \left( \frac{2 \pi i q}{nK} j \right)  \tilde{\Psi}_q (j) \nonumber \\
\Phi(j) = \exp \left( \frac{2 \pi i q}{nK} j \right)  \tilde{\Phi}_q (j) \\  
\eta(j) = \exp \left( \frac{2 \pi i q}{nK} j \right)  \tilde{\eta}_q (j) \nonumber 
\end{eqnarray}
for $1 \leq q \leq Kn$ and $\tilde{\Psi}_q, \tilde{\Phi}_q,\tilde{\eta}_q$
are assumed to be $n$-periodic. We will show that the vector $\mathbf{e}$ is an eigenvector of 
$D_{\mathbf{\hat{v}}^{(K)}}$ whenever the vector
\begin{equation} \label{mathi}
\mathbf{i} = (\tilde{\Psi}(j), \tilde{\Phi}(j), \tilde{\eta}(j) )_{j = 1}^n
\end{equation}
is an eigenvector of a particular  $3n \times 3n$ matrix
derived from $D_{\mathbf{\hat{v}}^{(K)}}F$.
 A direct calculation leads to:
\begin{align}\label{bloch}
\left[D_{\mathbf{\hat{v}}^{(K)}}F\right] \mathbf{e} \, (3j) = &
e^{ \frac{2 \pi i q}{nK} j }  ( d e^{ \frac{2 \pi i q}{nK} }
 \tilde{\Psi}_q (j+1 \!\!\!\!\!\mod n) + \nonumber 
de^{- \frac{2 \pi i q}{nK} } \tilde{\Psi}_q (j-1 \!\!\!\!\!\mod n) - \\
& 2 \tilde{\Psi}_q (j) + v_1(j) \tilde{\Psi}_q (j) + w_1(j) \tilde{\Phi}_q (j) +
u_1(j) \tilde{\eta}_q (j) ) \nonumber \\
\left[D_{\mathbf{\hat{v}}^{(K)}}F\right] \mathbf{e} \, (3j+1)= &
e^{ \frac{2 \pi i q}{nK} j } (v_2(j) \tilde{\Psi}_q (j) + w_2(j) \tilde{\Phi}_q (j) +
u_2(j) \tilde{\eta}_q (j))\\
\left[D_{\mathbf{\hat{v}}^{(K)}}F\right] \mathbf{e} \, (3j+2)= & 
e^{ \frac{2 \pi i q}{nK} j } (v_3(j) \tilde{\Psi}_q (j) + w_3(j) \tilde{\Phi}_q (j) +
u_3(j) \tilde{\eta}_q (j)) \nonumber
\end{align}  
for $1 \leq j \leq Kn$.  Here $v_i(j), w_i(j), u_i(j)$ are given by equation (\ref{linlorenz}) as
\begin{align*}
\begin{bmatrix}
v_1(j) &  w_1(j) & u_1(j)\\ 
v_2(j) &  w_2(j) & u_2(j)\\
v_3(j) &  w_3(j) & u_3(j) 
\end{bmatrix}  \quad = \quad &
\begin{bmatrix}
-\sigma & \sigma & 0 \\
r - \hat{z}_{j \!\!\!\!\! \mod n} & -1 & - \hat{x}_{j \!\!\!\!\!\mod n} \\
\hat{y}_{j \!\!\!\!\!\mod n} & \hat{x}_{j \!\!\!\!\!\mod n} & - b 
\end{bmatrix}
\end{align*}
and are thus $n$-periodic.

We want to check when $\mathbf{e}$ is an eigenvector of $D_{\mathbf{\hat{v}}^{(K)}}F$
and so we set $D_{\mathbf{\hat{v}}^{(K)}}F \mathbf{e} = \lambda \mathbf{e}$.  Using the
expressions obtained in  
(\ref{bloch}) we get
\begin{equation}\label{bloch2}
\begin{split}
 & \exp \left( \frac{2 \pi i q}{nK} j \right) \biggl( d e^{ \frac{2 \pi i q}{nK} }
 \tilde{\Psi}_q (j+1 \!\!\!\!\!\mod n) +
d e^{- \frac{2 \pi i q}{nK} } \tilde{\Psi}_q (j-1 \!\!\!\!\!\mod n) - \\
& 2 \tilde{\Psi}_q (j) + v_1(j) \tilde{\Psi}_q (j) + w_1(j) \tilde{\Phi}_q (j) +
u_1(j) \tilde{\eta}_q (j), \;\; 
v_2(j) \tilde{\Psi}_q (j) + w_2(j) \tilde{\Phi}_q (j) +
u_2(j) \tilde{\eta}_q (j), \\
& v_3(j) \tilde{\Psi}_q (j) + w_3(j) \tilde{\Phi}_q (j) +
u_3(j) \tilde{\eta}_q (j) \biggr)_{j = 1}^{Kn} = 
\lambda \exp \left( \frac{2 \pi i q}{nK} j  \right) \biggl(
\tilde{\Psi}_q(j), 
\tilde{\Phi}_q(j), \tilde{\eta}_q(j) \biggr)_{j = 1}^{Kn}.
\end{split}
\end{equation}

Since all the functions on the right hand side of (\ref{bloch2}) are 
$n$-periodic we can express equation (\ref{bloch2}) as $\mathcal{D}(q)
\mathbf{i} = \lambda \mathbf{i}$ where $\mathbf{i}$ is defined in
equation (\ref{mathi}) and
\begin{equation} 
\mathcal{D}(q) = \begin{bmatrix}\label{dq}
D_{\mathbf{\hat{v}_1}}f - 2\Gamma & E(q)^+ & 0 & \dots & 0 & E(q)^- \\
E(q)^-  & D_{\mathbf{\hat{v}_2}}f  - 2\Gamma& E(q)^+ & \dots & 0 & 0 \\
\hdotsfor[3]{6}                                   \\
E(q)^+   & 0 & 0 & \dots &  E(q)^- & D_{\mathbf{\hat{v}_n}}f -2\Gamma 
\end{bmatrix}
\end{equation}
with the matrices $E(q)^{\pm}$ defined as:
\begin{equation}
E(q)^{\pm} = \begin{bmatrix}
d e^{ \pm \frac{2 \pi i q}{nK}} & 0 & 0 \\
0 & 0 & 0 \\
0 & 0 & 0 
\end{bmatrix}
\end{equation}
This implies that the  eigenvalues of $D_{\mathbf{\hat{v}}^{(K)}}F$
are the same as the eigenvalues of the $3n \times 3n$ matrices $\mathcal{D}(q)$ for
$1 \leq q \leq Kn$. The $3n \times 3n$ matrix $\mathcal{D}(q)$ is  easier to handle than 
the $3Kn \times 3Kn$ matrix $D_{\mathbf{\hat{v}}^{(K)}}F$.  

To conclude that any $K$-multiple of a stable fixed state 
$\mathbf{v}_0$ of a ring
of $n$ oscillator is stable it is sufficient to show that the eigenvalues
of the matrix $\mathcal{D}(q)$ have negative real part for all values of 
$q$ and $K$.
To simplify the calculations we can replace the argument $q \in \Bbb{Z}$
by a continuous parameter $t \in \Bbb{R}$ by replacing the matrices 
$E(q)^{\pm}$ with the matrices
\begin{equation*}
E_t^{\pm} = \begin{bmatrix}
d e^{ \pm it} & 0 & 0 \\
0 & 0 & 0 \\
0 & 0 & 0 
\end{bmatrix}
\end{equation*}
in the expression for $\mathcal{D}$.  If we can show that $\mathcal{D}(t)$ 
has eigenvalues with negative real part for all $t \in \Bbb{R}$ then the same
is true for $\mathcal{D}(q)$ for any values of $K$ and $q$ in $\Bbb{Z}$.  

The following lemma about the class of matrices of type (\ref{dq})
will be used in the next section.

\begin{lemma} \label{char}
Let $E(t)^{\pm}$ be as in the previous lemma. 
The characteristic polynomial of a $Km \times Km$  matrix
\begin{equation*}
B = \begin{bmatrix}
M_1  & E(t)^+ & 0 & \dots & 0& 0 & E(t)^- \\
E(t)^-  & M_2 & E(t)^+ & \dots 0& & 0 & 0 \\
\hdotsfor[3]{6}  \\
 0 & 0 & 0 & \dots &  E(t)^- & M_{K-1} &  E(t)^+   \\
E(t)^+   & 0 & 0 & \dots & 0 & E(t)^- & M_K 
\end{bmatrix}
\end{equation*}
is of the form
\begin{equation}\label{charpoly}
p_M(t, \lambda) = \sum_{i=0}^{(K-1)m} (\alpha_i + \beta_i \cos Kt)  \lambda^i 
+ \sum_{i=(K-1)m+1}^{Km} \alpha_i \lambda^i .
\end{equation}
\end{lemma} 

A proof of this lemma is given in the appendix.

\section{An Example} \label{example}

This section gives a computer assisted proof that any $K$-multiple
of a particular stable fixed state in a ring of four Lorenz system is
stable.  The proof involves the use of interval arithmetic and the 
following theorems  about the convergence of the
Newton-Raphson-Kantorovich method.  The computations were performed
using Mathematica's implementation of interval arithmetic.  

\begin{theorem} \label{newton}
Let $f(z)$ be a complex analytic function and
assume $f(z_0) f'(z_0) \neq 0$ for some $z_0$.  Define $h_0 = -f(z_0)/f'(z_0)$,
the disc $K_0 = \{z : |z - z_0| \leq |h_0|\}$ and 
$M = \max_{K_0} |f''(z)|$.  If $2|h_0| M \leq |f'(z_0)|$ then
there is exactly one root of $f$ in the closed disc $K_0$.
\end{theorem}

\begin{theorem}[Kantorovich's Convergence Theorem] \label{kantorovich}
Given a twice differentiable 
function $f : \Bbb{R}^n \rightarrow \Bbb{R}^n$ which is nonsingular
at a point $x_0 = ( x_1^0, x_2^0, \dots, x_n^0 )$ let 
 $[\Gamma_{ik}] = [\frac{\partial f_i}{\partial x_k}(x_0)]^{-1}$. 
 Let $A, B, C$ be positive real numbers such that
\begin{align}
\max_i \sum_{k=1}^{n} |\Gamma_{ik}| \leq A \qquad &
\max_i \sum_{k=1}^{n} |\Gamma_{ik} f_k(x_0)| \leq B \qquad 
C \leq \frac{1}{2 A B}.
\end{align}
Define the region $R = \{x \in \Bbb{R}^n : \max_i |x_i - x_i^0|
\leq (AC)^{-1}(1 - \sqrt{1-2ABC}) \}$.  If
\begin{align}
\max_{x \in R} \sum_{j=1}^{n} \sum_{k=1}^{n} \left| 
\frac{\partial^2 f_i}{ \partial x_j \partial x_k}
\right| \leq C \qquad & i = 1, \dots,n
\end{align}
then the equation $f(x) = 0$ has a solution in $R$.
\end{theorem}

The argument will proceed as follows: First a steady state of the ring
is found
using interval arithmetic.  Next it is shown that the characteristic polynomial
$p_{\mathcal{D}}(\lambda, t)$ of the matrix $\mathcal{D}(q)$ in
(\ref{dq}) corresponding to this steady state has roots with
negative real part for $t = \frac{\pi}{4}$.  Finally it is proved that 
$p_{\mathcal{D}}(\lambda, t)$ does not have roots on the imaginary
axis for any $t \in \mathbb{R}$.  
Since the roots of a polynomial depend continuously on its
coefficients this implies that the roots of $p_{\mathcal{D}}(\lambda,
t)$ cannot cross into right half of $\mathbb{C}$.  By the results of
section \ref{stabilityofsteady} this implies that all $K$-multiples of
the fixed state under consideration must be stable.  

In the following intervals of numbers will be denoted with overbars to distinguish
them from real numbers, and make the notation less cumbersome.  
For instance $\bar{\lambda}$ denotes 
an interval and the interval  $[3.16524, 3.16533]$ is denoted  $\overline{3.165}$.

Numerical investigations with the program XPP 
show that a stable state in a chain of four
oscillators coupled with $d=1.5$ occurs close to
$ \hat{\mathbf{x}}= (\hat{x}_1,\hat{x}_2, \hat{x}_3, \hat{x}_4 )$ =  
(-6.4114408, 7.3696656, 8.1984129, 7.3696656). Notice that 
the orbit of (\ref{lorenzfixed}) describing this state is not symmetric 
in the sense of proposition \ref{devaney} although it is mapped into
itself under $R$.  This numerically obtained solution is used as an
initial guess in Mathematica's implementation of Newton's method
to determine the roots $x^i_a$ of the set of equations
\begin{equation} \label{stachain}
\sigma(x_i - \frac{\beta r x_i}{\beta+ (x_i)^2}) - 1.5 ( x_{(i-1)
\!\!\!\!\!\mod 4 +1} - 2 x_i + x_{(i+1)
\!\!\!\!\!\mod 4 +1}) \qquad i = 1,2,3,4.
\end{equation}
Although Mathematica can be instructed to find roots to an arbitrary
precision, so far only floating point arithmetic was used so
that none of the results are rigorous yet.  

At this point all the quantities in the computations are redefined as
intervals rather than floating point numbers and using interval
arithmetic and theorem
\ref{kantorovich} we find an interval around
each $x^i_a$ in which the roots of equation (\ref{stachain})
must lie.  These bounds are now rigorous.   

In section \ref{stabilityofsteady}  it was shown that 
a fixed state and all of its $K$-multiples are stable if 
the polynomial $p_{\mathcal{D}}(\lambda, t)$ corresponding to that
state
has roots with
negative real part for all $t \in [0, \pi/2]$.  Using interval
arithmetic it is shown that
\begin{equation}
\begin{split}
\bar{p}_{\mathcal{D}}(\lambda, t)= &
\overline{3.61524}\times{{10}^{12}} - \overline{5.02753}\times10^7 \cos 4t +  
(\overline{9.49583}\times10^{11} - \overline{1.33199}\times10^7 \cos 4t ) \,\lambda + \\
& (\overline{2.67802}\times10^{11} - \overline{4.94338}\times10^6 \cos 4t ) \,\lambda^2+  
(\overline{4.90382}\times10^{10} - \overline{771725} \cos 4t) \,\lambda^3 + \\ 
& (\overline{7.551}\times10^9 - \overline{144880} \cos 4t) \, \lambda^4 + 
(\overline{9.81388}\times10^8 - \overline{13673.3} \cos 4t) \lambda^5 + \\
& (\overline{1.03403}\times10^8 - \overline{1560.66} \cos 4t) \lambda^6 + 
(\overline{9.39525}\times10^6 - \overline{74.25} \cos 4t) \lambda^7 + \\
& (\overline{703005} - \overline{5.0625} \cos 4t) \lambda^8 + 
\overline{42026.7}\,\lambda^9 + \overline{2023.92}\,\lambda^{10} + 
  \overline{66.6667}\lambda^{11} + \lambda^{12}
\end{split}
\end{equation}
for the fixed state under consideration. 

In the next step the roots of the polynomial 
$\bar{p}_{\mathcal{D}}(\lambda, \frac{\pi}{2})$ are found.
To employ Mathematica's implementation of Newton's method we need 
a polynomial with coefficients that are floating point numbers rather
than intervals. Floating point numbers inside the intervals which
determine the coefficients of the 
polynomial $\bar{p}_{\mathcal{D}}(\lambda, \frac{\pi}{2})$ can be chosen 
to define a polynomial
$p^a_{\mathcal{D}}(\lambda, \frac{\pi}{2})$ 
which approximates $p_{\mathcal{D}}(\lambda, \frac{\pi}{2})$.
The roots of $\{r^a_i\}_{i=1}^{12}$ of 
$p^a_{\mathcal{D}}(\lambda,\frac{\pi}{2} )$ are now found using
Newton's method.  

The complex intervals\footnote{There are several
ways in which complex intervals can be defined.  In this case we use regions of the
form $\bar{z} = \bar{x} + i \bar{y}$ where $\bar{x}$ and $\bar{y}$ are real intervals.
These are simply rectangular regions in $\Bbb{C}$.} containing the roots of  
$\bar{p}_{\mathcal{D}}(\lambda,\frac{\pi}{2} )$ are found by
using theorem \ref{newton}.  We 
set $z_0$ equal to complex intervals around $r^a_i$ and use complex
interval arithmetic to find the radius $K_0$ and check the conditions
of the theorem for each root  $r^a_i$. 
The 4 real roots and 4 complex pairs
obtained by this procedure are given in the table below.
 
\vspace{5mm}
\centerline{
\begin{tabular}{||c|c||} \hline
$\overline{-18.481}$ & $\overline{-0.2463} \pm i \overline{9.769}$  \\ \hline
$\overline{-17.173}$ & $\overline{-0.1961} \pm i \overline{9.317}$ \\ \hline
$\overline{-15.437}$ & $\overline{-0.1345} \pm i \overline{9.164}$ \\ \hline
$\overline{-14.238}$ & $\overline{-0.09054} \pm i \overline{8.622}$ \\ \hline
\end{tabular}
}
\vspace{5mm}

Since interval arithmetic was used in these calculations, these
estimates are now rigorous. 
The remainder of the argument shows that these roots will not cross the imaginary
axis as $t$ varies.

By lemma \ref{char} the characteristic polynomial $p_{\mathcal{D}}(\lambda, t)$ takes
the following form when evaluated on the imaginary axis:
\begin{equation}
\begin{split}
 p_{\mathcal{D}}(i\mu, t) & = \sum_{j=0, j \text{ even}}^{12} (\alpha_j + \beta_j \cos 4t)
\mu^j (-1)^{\frac{j}{2}} + i \sum_{j=1, j \text{ odd}}^{11} (\alpha_j + \beta_j \cos 4t)
\mu^j (-1)^{\frac{j-1}{2}} = \\
& = p_{\mathcal{D}}^{\text{R}}(\mu, t) + i p_{\mathcal{D}}^{\text{I}}(\mu, t) \nonumber
\end{split}
\end{equation}
Since the roots of $p_{\mathcal{D}}(\lambda, t)$ depend continuously on the parameter $t$,
if $p_{\mathcal{D}}(\lambda, t)$ has roots with positive real part for some $t_1$ then
there must exist a $t_0$ such that $p_{\mathcal{D}}(\lambda, t_0)$ has a root on the
imaginary axis.  In other words there must exist a $\mu_0$ such that
\begin{equation} \label{inequality}
p_{\mathcal{D}}^{\text{R}}(\mu_0, t_0) =p_{\mathcal{D}}^{\text{I}}(\mu_0, t_0)=0.
\end{equation}
We will use interval arithmetic to show that this cannot happen. The polynomial
$p_{\mathcal{D}}(\lambda, t)$ is split  into a real and complex part to avoid
using complex interval arithmetic in the numerical calculations since complex
interval arithmetic leads to much rougher estimates than real interval arithmetic.

By Gre\v{s}gorin's theorem the eigenvalues of the matrix $\mathcal{D}(t)$ lie
in the union of discs $C_i= \{ z: |z-\mathcal{D}_{ii}| < R_i \}$ where 
$\mathcal{D}_{ij}$ are the entries in the matrix $\mathcal{D}(t)$ and
$R_i = \sum_{j=1, j \neq i}^{12} |\mathcal{D}_{ij}|$. A direct computation
shows that the intersection of $\cup_{i=1}^{12} C_i$ with the imaginary
axis is contained in the interval $[-17 i, 17i]$ and it is therefore
sufficient to show that $p_{\mathcal{D}}^{\text{R}}(\mu, t_0)$ and
$p_{\mathcal{D}}^{\text{I}}(\mu, t_0)$ are not zero simultaneously for 
any value of $t$ and $\mu \in [-17,17]$.

Since the coefficients of $\bar{p}_{\mathcal{D}}(\lambda,
t)$ are $\frac{\pi}{2}$ periodic in $t$, it is sufficient to show that
(\ref{inequality}) is not satisfied for any $\lambda \in [-17,17]$ and
$t \in [0, \frac{\pi}{2}]$.  This is shown by subdividing these intervals into 
a sufficient number of subintervals $\bar{\mu}_n$ and $\bar{t}_m$ so that 
$[-17,17] \subset \cup_{i=1}^{N} \bar{\mu}_n$ and 
$[0,\frac{\pi}{2}] \subset \cup_{m=1}^{M} \bar{t}_m$ with the property that
the intervals 
$\bar{p}_{\mathcal{D}}^{\text{R}}(\bar{\mu}_n, \bar{t}_m)$ and
$\bar{p}_{\mathcal{D}}^{\text{I}}(\bar{\mu}_n, \bar{t}_m)$ do not
contain zero simultaneously for any given pair of subintervals
$\bar{\mu}_n$ and $\bar{t}_m$.
Therefore the roots of $p_{\mathcal{D}}(\lambda, t)$ stay in the left half plane
for all $t \in \mathbb{R}$ and all 
$K$-multiples of the fixed state under consideration are also stable. 

\begin{figure} 
\centering\epsfig{figure=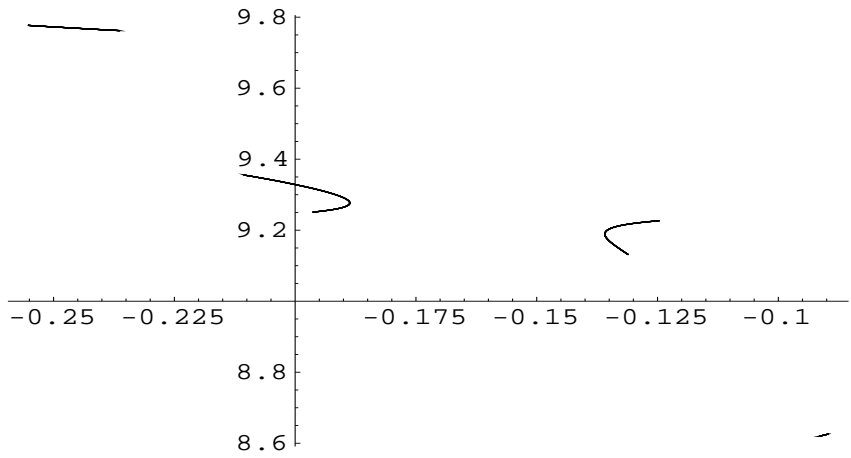, width=.8\linewidth, height = 60mm}\\
\caption{ A numerical computation of the paths that the 8 complex
eigenvalues with smallest real part trace out as $t$ varies.  Only
one eigenvalue in the complex conjugate pair is shown. Notice the
small trace of the eigenvalue with the least negative real part.} \label{eigenvalues}  
\end{figure}

The paths of the roots of $p_{\mathcal{D}}(\lambda, t)$ as $t$ is
varied are shown in figure \ref{eigenvalues}.

The \emph{Mathematica} code used in these calculations is available at

\emph{http://math.bu.edu/people/josic/research/code}.

{\bf Acknowledgment:} The authors research was supported in part by
NSF grant DMS-9803164.  K.J. would like to thank the Department of
Mathematics at Boston University for its hospitality while he
conducted his research. The authors also thank J.P. Eckmann for 
useful discussions concerning this problem.

\appendix

\section{Proof of Lemma \ref{char}}

\begin{sublemma}\label{prelim}
Assume $A$ is an $(n+m) \times (n+m)$ matrix of the form
\begin{equation}
A = \begin{bmatrix}
B   & D^+ & 0 & \dots & 0 \\
D^- &     &   &       &   \\
0   &     &\mbox{\LARGE {C}} &       &   \\
\vdots&   &   &       &   \\
0   &     &   &       &   \\
\end{bmatrix}
\end{equation}
where $m>n$, $B$ and $C$ are arbitrary $n \times n$, respectively $m \times m$
matrices and $D^{\pm}$ are $n \times n$ matrices of the form 
\begin{equation}
D^{\pm} = \begin{bmatrix}
d_{\pm} & 0 & \dots & 0 \\
0 & 0 & \dots& 0 \\
\hdotsfor[3]{4}\\
0 & 0& \dots & 0 
\end{bmatrix}.
\end{equation}
Writing the determinant of $A$ as
\begin{equation} \label{det}
\det A = \sum_{\sigma} \mbox{sgn}\; \sigma \prod_{i=1}^{m+n} a_{i, \sigma(i)}
\end{equation}
then if $a_{1,\sigma(1)} = a_{1, n+1} = d_+$ then
\begin{equation} \label{product}
\prod_{i=2}^{M+n} a_{i, \sigma(i)} \neq 0
\end{equation}
only if $a_{n+1,\sigma(n+1)} = a_{n+1, 1} = d_-$.
\end{sublemma}

This lemma simply states that the only nonzero products in (\ref{det}) containing
$d_+$ as a factor necessarily contain $d_-$ as a factor.  Notice that since
we can interchange rows and columns when calculating determinants the opposite is
also true: nonzero products containing $d_-$ as a factor necessarily contain
$d_+$ as a factor. 

{\bf Proof:} Write the matrix $A$ as
\begin{equation}
A = \begin{bmatrix}
b_{1,1}  & b_{1,2} & \dots & b_{1,n}   & d_+ & 0 & \dots  \\
b_{2,1}  & b_{2,2} & \dots & b_{2,n}   & 0  & 0 & \dots  \\
\hdotsfor[3]{7}\\
d_-      & 0       & \dots & 0          &    &   &         \\
0        & 0       & \dots & 0          &    &   &         \\
\vdots   & \vdots  & \vdots & \vdots     &    & \mbox{\LARGE{C}} &         \\
0        & 0       & 0     & 0          &    &   &       \\
\end{bmatrix}.
\end{equation}
If $a_{n+1,\sigma(n+1)} = a_{n+1, 1} = d_+$ and $\sigma(i) = 1$ for some $2 \leq i \leq n$ 
another $n-1$ nonzero factors in the product (\ref{product}) need to be
chosen from columns 2 through $n$.  However, barring rows 1 and $i$ which
were already used, there are only $n-2$ rows remaining which have nonzero
entries in these columns.  This shows that factors 
$a_{2, \sigma(2)}, a_{3, \sigma(3)}, \dots, a_{n, \sigma(n)}$ must come from
columns 2 through $n$.  Since the rows 1 through $n$ are now exhausted, the only nonzero
element remaining in column 1 is $d_-$.  This proves the lemma. $\diamondsuit$

\begin{sublemma} \label{mat}
The characteristic polynomial of the $Km \times Km$ matrix 
\begin{equation*}
A = \begin{bmatrix}
M_1  & E(t)^+ & 0 & \dots & 0 & 0 & 0 \\
E(t)^-  & M_2 & E(t)^+ & \dots & 0  & 0 & 0 \\
\hdotsfor[3]{6}  \\
0& 0 & 0 & \dots & E(t)^- & M_{K-1} &  E(t)^+ \\
0   & 0 & 0 & \dots &  0  & E(t)^- & M_K 
\end{bmatrix}
\end{equation*}
where $M_i$ are arbitrary $m \times m$ matrices and  $E(t)^{\pm}$ are $m \times m$ matrices defined as
\begin{equation*}
E(t)^{\pm} = \begin{bmatrix}
d e^{ \pm it} & 0 & \dots & 0 \\
0 & 0 & \dots& 0 \\
\hdotsfor[3]{4}\\
0 & 0& \dots & 0 
\end{bmatrix}
\end{equation*}
does not depend on $t$.
\end{sublemma}

{\bf Proof:} Define the matrix $C = A - \lambda I$.  
We need to show that the determinant of $C$ is independent of $t$.
The proof is by induction on $K$.  When $K=1$ the matrix $C$ is itself $t$
independent so the statement is trivial.  The cofactors of the entries $c_{1,j}$ for
$1 \leq j \leq m$ are of the form
\begin{equation}
\begin{bmatrix}
c_{2,1}  & \dots & \hat{c}_{2,j} & \dots & c_{1,m}  & 0 & \dots & 0 \\
\hdotsfor[3]{7}\\
c_{m,1}  & \dots & \hat{c}_{m,j} & \dots & c_{m,m}  & 0 & \dots & 0 \\
de^{-it}      & 0       & \dots & \dots          & 0   &   &   &      \\
0        & 0       & \dots & \dots          & 0   &   &   &      \\
\vdots   & \vdots  & \vdots & \vdots     &  \vdots  &  & \mbox{\LARGE{M}} &       \\
0        & 0       & 0     & 0          &  0  &   &   &    \\
\end{bmatrix}
\end{equation}
where the hat signifies the omission of that particular column.
Since the columns and rows can be interchanged when the determinant is computed lemma~\ref{prelim}
implies that any product in (\ref{det}) containing $de^{-it}$ as a factor also contains 
$c_{1,n+1}=0$ as a factor and is thus zero.  Lemma~\ref{prelim} also implies that
any nonzero product containing $c_{1,n+1}=de^{it}$ as a factor necessarily contains $c_{1,n+1}=de^{-it}$
as a factor and so does not depend on $t$ either.  Since this exhausts all possible cofactors 
of the nonzero elements of the first row the lemma is proved. $\diamondsuit$

Lemma \ref{char} can now be proved. We need to show that the
determinant of  $C = B - \lambda I$ is of the form (\ref{charpoly}).
Definition (\ref{det}) will again be used for the determinant. We now have three cases:

\emph{Case 1:} Assume that for $\sigma$ in the product in (\ref{det}) we have
$1 \leq \sigma(1) \leq m$.  Since the next $m-1$ rows have
zero entries in all but the first $m$ columns this particular product will be zero unless 
it contains $c_{j,1}$ as a factor for some $1 \leq j \leq m$.  This
excludes the possibility that either $c_{1, (K-1)m+1} = de^{-it}$ or 
$c_{ (K-1)m+1,1} =  de^{it}$ appear as factors.  If we fix
$\sigma(1), \dots, \sigma(m)$, the part of the sum (\ref{det}) corresponding
to all such permutations $\sigma$ is equal to
\begin{equation}
c_{1,\sigma(1)} \dots c_{m,\sigma(m)} 
\begin{vmatrix}
M_2  & E(t)^+ & 0 & \dots & 0& 0 & 0 \\
E(t)^- & M_3 & E(t)^+ & \dots 0& & 0 & 0 \\
\hdotsfor[3]{6}  \\
 0 & 0 & 0 & \dots &  E(t)^- & M_{K-1} &  E(t)^+   \\
0   & 0 & 0 & \dots & 0 & E(t)^- & M_K 
\end{vmatrix} 
\end{equation}
by cofactor expansion.  By sublemma~\ref{mat} these summands will not depend on $t$.

\emph{Case 2:} Assume that $\sigma(1) = m+1$ so that
the first factor in the product in (\ref{det}) is $c_{1,m+1}=de^{it}$.  Then the elements
$c_{2,\sigma(2)}, \dots, c_{m, \sigma(m)}$ must come from the matrix
\begin{equation*}
\tilde{C} = \begin{bmatrix}
c_{2, 2}  & \dots & c_{2, m}    \\
c_{3,2}         & \dots & c_{3, m}   \\
\hdotsfor[3]{3} \\
c_{m,2}         & \dots & c_{m, m}   
\end{bmatrix}
\end{equation*}
by an argument identical to the one given in the second paragraph of sublemma (\ref{prelim}). The only nonzero 
factors that can be chosen from column 1 are therefore $c_{m+1,\sigma(m+1)} = c_{m+1,1} = de^{-it}$ and
$c_{(K-1)m+1,\sigma(K-1)m+1} = c_{(K-1)m+1,1}= de^{it}$.  We therefore have the following two
subcases:

\emph{Case 2a:} If $c_{m+1,1} =de^{-it}$ is chosen it cancels the term 
$c_{1,m+1}=de^{it}$ that was chosen as the first factor in the product.
In this case we are again in the situation of
sublemma~\ref{mat} using Laplace expansion as in case 1, and so this product will not depend on $t$.

The contributions to the characteristic polynomial from Case 1 and 2a 
are time independent and therefore enter the term $\sum_{i=(K-1)m+1}^{Km} \alpha_i \lambda^i$
in (\ref{charpoly}).  The remainder of the lemma follows from induction.

\emph{Case 2b:} If the assumptions of Case 2 hold and 
$c_{(K-1)m+1,\sigma((K-1)m+1)}=c_{(K-1)m+1,1} = d e^{it}$
then if a permutation $\sigma$ leads to a nonzero product in (\ref{det})
it must satisfy $\sigma(jm+1)=(j+1)m+1$ for all $j < K-1$.  This simply means
that any nonzero product of this kind contains all the
entries $c_{jm+1,(j+1)m+1}=de^{it}$ as factors.  This assertion is proved as follows:  

The factor $c_{1,m+1}$ is contained in the product by assumption.
Assume that $\sigma(lm+1) = (l+1)m+1$ for all $l < j$.  It will
be shown that in this case $\sigma(jm+1) = (jm+1)m+1$ which will complete 
the induction argument.  

By assumption the $\sigma((j-1)m+1) = jm+1$.  The only 
nonzero elements in columns $jm+2, \dots (j+1)m$ come from the submatrix
\begin{equation*}
\begin{bmatrix}
c_{jm+2, jm+2}  & \dots & c_{jm+2, (j+1)m}    \\
c_{jm+3,jm+2}         & \dots & c_{jm+3, (j+1)m}   \\
\hdotsfor[3]{3} \\
c_{(j+1)m,jm+2}         & \dots & c_{(j+1)m, (j+1)m}   
\end{bmatrix}
\end{equation*}
and since an entry from row  $jm+1$ has already been chosen, the coefficients
$c_{jm+2, \sigma(jm+2)}, $ $\dots,$ $c_{(j+1)m, \sigma((j+1)m)}$ must all come from 
this submatrix.  The only nonzero entries remaining on row $jm+1$ are
$c_{jm+1, (j-1)m+1} = de^{-it}$ and $c_{jm+1, (j+1)m+1} = de^{it}$. 
By the induction hypothesis we can exclude the first possibility and
we are left to conclude that $c_{jm+1, (j+1)m+1} = de^{it}$ for any $1 < j < K$ is a coefficient
in any nonzero product in the case 2b.     

This shows that any permutation $\sigma$ of the type described in Case 2b contributes
a factor $\gamma_{\sigma} e^{Kit} r(\lambda)$ to the characteristic polynomial
(\ref{charpoly}). 

\emph{Case 3:} The only remaining nonzero entry in row 1 of matrix $C$ is $c_{1,(K-1)m+1} =de^{-it}$.
An argument identical to the one presented in case 2 shows that any permutation $\sigma$
leading to a nonzero summand in (\ref{det}) must satisfy either
$c_{(K-1)m+1, \sigma((K-1)m+1)} = c_{(K-1)m+1,1} = de^{it}$ or
$c_{(K-1)m+1, \sigma((K-1)m+1)} = c_{(K-1)m+1,(K-2)m+1} = de^{-it}$
which again leads to two subcases:

\emph{Case 3a:} If $c_{(K-1)m+1, \sigma((K-1)m+1)}= de^{it}$ the situation is
virtually identical to case 2a.  Laplace expansion and
sublemma \ref{mat} show that the contributions to the characteristic polynomial are 
independent of $t$.

\emph{Case 3b:} If $c_{(K-1)m+1, \sigma((K-1)m+1)}= de^{-it}$ the situation
is similar to case 2b.  A parallel argument shows that all the terms $de^{-it}$
in the matrix must be coefficients in any nonzero product.  Moreover due to
the properties of the matrix, to each product $\gamma_{\sigma} e^{Kit} r(\lambda)$
from case 2b 
corresponds a product $\gamma_{\sigma} e^{-Kit} r(\lambda)$ from this case.  Since
these are the only time dependent contributions to the characteristic polynomial the
lemma is proved. $\diamondsuit$

\pagebreak

\listoffigures

\pagebreak

\pagebreak

\pagebreak

\pagebreak

\pagebreak

\pagebreak

\pagebreak

\pagebreak

\pagebreak

\pagebreak

\pagebreak

\end{document}